\documentclass{lmcs}
\pdfoutput=1
\usepackage[utf8]{inputenc}

\usepackage{lastpage}
\lmcsdoi{22}{2}{9}
\lmcsheading{}{\pageref{LastPage}}{}{}%
{Nov.~22,~2024}{Apr.~21,~2026}{}


\usepackage{xspace}
\usepackage{amssymb}
\usepackage{amsthm}
\usepackage{xcolor,colortbl}
\usepackage{listings}
\usepackage{makecell}
\usepackage{varwidth}
\usepackage{multirow}
\usepackage[noend]{algorithmic}
\usepackage{hyperref}
\usepackage{cleveref}
\usepackage{booktabs}
\Crefname{thm}{Theorem}{Theorems}
\Crefname{thmC}{Theorem}{Theorems}
\Crefname{lem}{Lemma}{Lemmata}
\Crefname{prop}{Proposition}{Propositions}
\Crefname{cor}{Corollary}{Corollaries}
\Crefname{rem}{Remark}{Remarks}
\Crefname{defi}{Definition}{Definitions}
\Crefname{exa}{Example}{Examples}

\newcommand{\defeq}{\vcentcolon=}

\def\polylog{\operatorname{polylog}}

\newcommand{\sparseBMM}{\textsc{sparseBMM}}

\newcommand{\atoms}{\mathrm{atoms}}
\newcommand{\var}{\mathrm{vars}}

\newcommand{\free}{\mathrm{free}}

\newcommand{\powerset}{{\mathcal{P}}}

\newcommand{\ar}{\texttt{ar}}

\def \before{_{\mathsf{before}}}
\def \after{_{\mathsf{after}}}
\def \carry{_{\mathsf{carry}}}
\def \last{{\mathsf{last}}}

\def\neigh{\mathsf{N}}

\newcommand{\N}{\mathbb{N}}

\newcommand{\Z}{{\mathbb{R}}}

\def\e#1{\emph{#1}}

\def\angs#1{\mathord{\langle #1\rangle}}
\def\loglin{\mathrm{loglinear}}
\def\lin{\mathrm{linear}}
\def\linlog{\angs{\loglin,\log}}
\AtBeginDocument{
  \providecommand\BibTeX{{
    \normalfont B\kern-0.5em{\scshape i\kern-0.25em b}\kern-0.8em\TeX}}}

\newcommand{\datarule}{{\,:\!\!-\,}}

\usepackage{tabularx}
\usepackage{booktabs}

\newcommand{\hide}[1]{} 
\newcommand{\hidetwo}[2]{}

\def\scs{\mathord{\mathbf{S}}}
\def\zero{\mathord{\bar{0}}}
\def\one{\mathord{\bar{1}}}
\def\K{\mathbb{K}}

\def\set#1{\{#1\}}
\def\vals{\mathsf{Vals}}

\def\hom{\mathsf{Hom}}

\def\Q{\mathbb{Q}}
\def\Z{\mathbb{Z}}
\def\U{\mathcal{U}}
\def\V{\mathcal{V}}

\def\rel#1{\textsc{#1}}
\def\att#1{\textit{#1}}
\def\dla{\mathrel{{:}{-}}}
\def\aggcount{\mathsf{Count}}
\def\aggsum{\mathsf{Sum}}
\def\aggmin{\mathsf{Min}}
\def\aggmax{\mathsf{Max}}
\def\aggcountd{\mathsf{CountD}}
\def\aggavg{\mathsf{Avg}}

\newcommand{\minf}{\mathsf{min}}
\newcommand{\maxf}{\mathsf{max}}

\def\fd{_{\mathsf{fd}}}
\def\Qstartimes{Q_{\star\times}}
\def\Qaggcount{Q_{c}}
\def\pfree#1{#1_{|\free{(#1)}}}
\def\pfreeQp#1{#1_{|\free{(Q')}}}
\def\pfreeQ#1{#1_{|\free{(Q)}}}
\def\Qpfree{\pfree{Q}}
\def\Qtpfree{\pfree{Q'}}
\def\Qsffree{Q^\mathrm{sf}_{|\free{(Q)}}}
\def\Qsf{Q^\mathrm{sf}}

\def\hypo#1{$\mathsf{#1}$}
\def\sparseBMM{\hypo{SparseBMM}\xspace}
\def\HYPERCLIQUE{\hypo{HYPER\-CLIQUE}\xspace}
\def\THREESUM{\hypo{3SUM}\xspace}
\def\HSC{\hypo{HSC}\xspace}

\definecolor{Gray}{gray}{0.85}
\newcolumntype{g}{>{\columncolor{Gray}}c}

\def\full{\mathsf{full}}
\def\parcase#1#2{\bigskip\par\noindent\underline{\textbf{#1:}} \textit{#2}\,\,}

\def\cqstar{CQ$^\star$\xspace}
\def\cqstars{CQ$^\star$s\xspace}

\def\ACQ{AggCQ\xspace}
\def\ACQs{AggCQs\xspace}

\def\cntdomain{\mathord{\mathcal{X}}}

\theoremstyle{plain}

\begin{document}

\title{Direct Access for Answers to Conjunctive Queries with Aggregation}

\author[I.~Eldar]{Idan Eldar\lmcsorcid{0009-0002-1664-8680}}[a]
\author[N.~Carmeli]{Nofar Carmeli\lmcsorcid{0000-0003-0673-5510}}[b]
\author[B.~Kimelfeld]{Benny Kimelfeld\lmcsorcid{0000-0002-7156-1572}}[a]

\address{Technion -- Israel Institute of Technology, Haifa, Israel}
\email{idel@campus.technion.ac.il, bennyk@cs.technion.ac.il}

\address{Inria, LIRMM, University of Montpellier, CNRS, France}
\email{nofar.carmeli@inria.fr}

\begin{abstract}
We study the fine-grained complexity of conjunctive queries with grouping and aggregation. For common aggregate functions (e.g., min, max, count, sum), such a query can be phrased as an ordinary conjunctive query over a database annotated with a suitable commutative semiring.
We investigate the ability to evaluate such queries by constructing in loglinear time a data structure that provides logarithmic-time direct access to the answers ordered by a given lexicographic order. This task is nontrivial since the number of answers might be larger than loglinear in the size of the input, so the data structure needs to provide a compact representation of the space of answers. In the absence of aggregation and annotation, past research established a sufficient tractability condition on queries and orders. For queries without self-joins, this condition is not just sufficient, but also necessary (under conventional lower-bound assumptions in fine-grained complexity). 

We show that all past results continue to hold for annotated databases, assuming that the annotation itself does not participate in the lexicographic order. Yet, past algorithms do not apply to the count-distinct aggregation, which has no efficient representation as a commutative semiring; for this aggregation, we establish the corresponding tractability condition.
We then show how the complexity of the problem changes when we include the aggregate and annotation value in the order. We also study the impact of having all relations but one annotated by the multiplicative identity (one), as happens when we translate aggregate queries into semiring annotations, and having a semiring with an idempotent addition, such as the case of min, max, and count-distinct over a logarithmic-size domain.
\end{abstract}

\maketitle

\section{Introduction}
Consider a query $Q$ that may have a large number of answers, say cubic in the number of tuples of the input database $D$. By answering $Q$ via \e{direct access}, we avoid the materialization of the list of answers, and instead, construct a compact data structure $S$ that supports random access: given an index $i$, retrieve the $i$th answer. Hence, direct access evaluation for a query $Q$ consists of two algorithms, one for the structure construction (with the input $D$), called \e{preprocessing}, and one for fast access to the answers (with the input $S$ and $i$). 
This task is nontrivial when $S$ is considerably cheaper to construct than $Q(D)$. 
Similarly to past work on direct access~\cite{carmeli2023tractable}, we adopt the tractability requirement of linear or quasi-linear time to construct $S$, and logarithmic time per access.
Hence, up to a poly-logarithmic factor, the required construction time is what it takes to read the database (i.e., linear time), and the access time is constant.
The structure $S$ can be viewed as a compact representation of $Q(D)$, in the general sense of Factorized Databases~\cite{DBLP:journals/sigmod/OlteanuS16}, since its size is necessarily quasi-linear and it provides fast access.

Direct access solutions have been devised for Conjunctive Queries (CQs), first as a way to establish algorithms for enumerating the answers with linear preprocessing time and constant delay~\cite{braultbaron:tel-01081392} (and FO queries with restrictions on the database~\cite{bagan:hal-00221730}); the preprocessing phase constructs $S$, and the enumeration phase retrieves the answers by accessing $S$ with increasing indices $i$. Later, direct access had a more crucial role within the task of enumerating the answers in a uniformly random order~\cite{DBLP:journals/tods/CarmeliZBCKS22}. As a notion of query evaluation, direct access is interesting in its own right, since we can view $S$ itself as the ``result'' of the query in the case where array-like access is sufficient for downstream processing (e.g., to produce a sample of answers, to return answers by pages, to answer $q$-quantile queries, etc.).
But then $S$ has the benefit that it is considerably smaller and faster to produce than the materialized set of answers.
Indeed, recent work has studied the complexity of direct access independently (regardless of any enumeration context)~\cite{bringmann2025tight}, and specifically studied which \e{orders} over the answers allow for such evaluation~\cite{carmeli2023tractable}. In this paper, we continue with the line of work by Carmeli et al.~\cite{carmeli2023tractable} and investigate the ability to support query evaluation via direct access for \e{aggregate queries}, while focusing on lexicographic orderings of answers.

For illustration, consider the following example, inspired by the FIFA World Cup. Suppose that we have a database of players of teams (countries), sponsors of teams, and goals scored in different games. Specifically, we have three relations:
$\rel{Teams}(\att{player},\att{country})$,
$\rel{Sponsors}(\att{org},\att{country})$, 
and
$\rel{Goals}(\att{game},\att{player},\att{time})$. The following CQ finds times when sponsors got exposure due to goals of supported teams:
\[Q_1(c,o,p,t)\dla 
\rel{Teams}(p,c),
\rel{Sponsors}(o,c), 
\rel{Goals}(g,p,t)\]
Suppose also that we would like the answers to be ordered lexicographically by their order in the head: first by $c$ (country), then by $o$ (organization), then by $p$ (player), and lastly by $t$ (time).
Note that $o$, $c$, $p$ and $t$ are \e{free} variables and $g$ is an \e{existential} variable.
Carmeli et al.~\cite{carmeli2023tractable}
studied the ability to evaluate such ordered queries with direct access. In the case of $Q_1$, the results of Carmeli et al.~show that there is an efficient direct access evaluation (since the query is free-connex and there is no ``disruptive trio''). 

Now, suppose that we would like to count the goals per sponsorship and player. In standard CQ notation (e.g., Cohen et al.~\cite{10.1145/1219092.1219093}), we can phrase this query as follows.
\[Q_2(c,o,p,\aggcount())\dla \rel{Teams}(p,c),
\rel{Sponsors}(o,c), 
\rel{Goals}(g,p,t)\]
Here, the free variables $c$, $o$, and $p$ are treated as the \e{grouping variables}, where each combination of values defines a group of tuples over $(c,o,p,g,t)$ and $\aggcount()$ simply counts the tuples in the group. Again, we would like to answer this query via direct access. This introduces two challenges. The first challenge is \e{aggregate construction}: when we access a tuple using $S$, the aggregate value should be quickly produced as well. 
The second challenge is \e{ordering by aggregation}: how can we incorporate the aggregation in the lexicographic order of the answers if so desired by the query? As an example, we may wish to order the answers first by $c$, then by $\aggcount()$, and then by $o$ and $p$; in this case, we would phrase the head accordingly as $Q_2(c,\aggcount(),o,p)$. 

As previously done in the context of algorithms for aggregate queries~\cite{DBLP:journals/vldb/ReS09,DBLP:journals/tods/KhamisCMNNOS20}, we also study
a semiring alternative to the above formalism of aggregate queries. 
Specifically, we can adopt the well-known framework of \e{provenance semiring} of Green, Karvounarakis, and Tannen~\cite{10.1145/1265530.1265535} and phrase the query as an ordinary CQ with the annotation carrying the aggregate value (e.g., the number of goals in our example). To reason about random-access evaluation, we found it more elegant, general, and insightful to support CQs over annotated databases rather than SQL-like aggregate functions. For illustration, we can phrase the above aggregate query $Q_2$ as the following CQ $Q_3$, but
for a database that is annotated using a specific commutative semiring.
\[Q_3(c,o,p)\dla 
\rel{Teams}(p,c),
\rel{Sponsors}(o,c), 
\rel{Goals}(g,p,t)
\]
 In a nutshell (the formal definition is in Section~\ref{sec:prelim}), the idea is that each tuple is annotated with an element of the semiring, the annotation of each tuple in the group is the product of the participating tuple annotations, and the annotation of the whole group is the sum of all tuple annotations in the group's tuples. In the case of our example with $Q_3$, we use the counting semiring $(\N,+,\cdot,0,1)$, 
 and each tuple is annotated simply with the number $1$. We can use different semirings and annotations to compute different aggregate functions like sum, min, and max. Here again, we have challenges analogous to the aggregate case: \e{annotation construction} and \e{ordering by annotation}. The previous example becomes ordering by $c$, then by the annotation, and then by $o$ and $p$. Notationally, we specify the annotation position by the symbol $\star$ and phrase the query as 
$Q_3(c,\star,o,p)\dla 
\rel{Teams}(p,c),
\rel{Sponsors}(o,c), 
\rel{Goals}(g,p,t)$. We refer to such a query as a \cqstar.

In this paper, we study queries in both formalisms---CQs enhanced with aggregate functions and ordinary \cqstars over annotated databases. We usually devise algorithms and upper bounds on general commutative semirings (possibly with additional conditions), as positive results carry over to the aggregate formalism, and we prove cases of specific intractable queries with specific aggregate functions over ordinary (non-annotated) databases. 

Our analysis is done in two parts. In Section~\ref{sec:outside}, we study the case where the annotation or aggregation is \e{not} a part of the lexicographic order; we show that under reasonable assumptions about the complexity of the semiring operations, the known dichotomy for ordinary databases~\cite{carmeli2023tractable} continues to hold in the presence of annotation (hence, we can efficiently solve the aforementioned first challenge, namely annotation construction). We conclude the analogous tractability frontier for the common aggregate functions (count, sum, min, max, average). A notable exception is the count-distinct aggregation, which cannot be expressed efficiently as a semiring annotation; we show that the class of tractable queries for count-distinct is indeed more restricted, and we establish the precise tractability condition (dichotomy) for this aggregation.

In Section~\ref{sec:inside}, we study the ability to include the annotation or aggregation in nontrivial positions within the lexicographic order (i.e., the second challenge). We establish a sufficient tractability condition for \cqstars. Moreover, we prove that this tractability condition is also necessary (under standard assumptions in fine-grained complexity theory) for commonly studied semirings, when the \cqstar has no self-joins. We also investigate special cases of aggregate CQs where the complexity is not resolved 
by the dichotomy on \cqstars. Interestingly, the translation to \cqstars of CQs with the aforementioned aggregate functions has a special property: in all relations but one, the annotation is constantly the multiplicative identity (i.e., the ``one'' element). We refer to annotated databases with this property as \e{locally annotated}. In Section~\ref{sec:locally-annotated}, we examine the implication of local annotation on the complexity of direct access, that is, whether this property can be utilized to establish tractable queries that are hard in the absence of this property, and to what extent. We answer the first question affirmatively. Moreover, we establish a dichotomy similarly to the previous ones (i.e., the hardness side assumes that there are no self-joins) for the class of full \cqstars. We also study the implication of local annotation in conjunction of another property, namely that the addition operation is \e{idempotent}. This covers the min and max aggregations, as well as count-distinct in the case of a small (logarithmic-size) domain of counted entities. We establish the corresponding dichotomy for the class of \cqstars (with projection). 

As closely related work, we note that Keppeler~\cite{DBLP:phd/dnb/Keppeler20} has proposed algorithms for direct access with aggregation, where he considers the class of q-hierarchical CQs (that support efficient updates of the direct-access structure~\cite{DBLP:conf/pods/BerkholzKS17}). Yet, he does not discuss the dependence of tractability on the lexicographic order. Moreover, our positive results (e.g., \Cref{thm:annotated-dichotomy,thm:dichotomy-general-annotation}) apply to the class of free-connex CQs, which is more general than that of the q-hierarchical CQs.

An abridged version of this manuscript appeared in a conference proceedings~\cite{DBLP:conf/icdt/EldarCK24}. Compared to the conference version, this manuscript has significant additions and generalizations. First, we include here the full proofs of all results. Second, we generalized two results into full dichotomy theorems: the lower bound for a specific count-distinct query~\cite[Theorem~8]{DBLP:conf/icdt/EldarCK24}
is generalized to the dichotomy of \Cref{thm:countd-dichotomy} here, 
and the tractability condition for annotated databases (with the computed annotation taking part of the order)~\cite[Theorem~12] {DBLP:conf/icdt/EldarCK24} is 
now \Cref{cor:general-annotations-order-tractability} of the new
\Cref{thm:dichotomy-general-annotation} here (that extends the tractability criterion and shows its necessity). Third, we reorganized the results and proofs by providing new concepts (e.g., the restriction $Q_{|V}$ of a query $Q$ to a subset $V$ of its variables) and machinery (e.g., \Cref{lem:reduction} and \Cref{thm:existential-removal} that give general reducibility conditions
that are used throughout the article and are of interest independently of this work). Fourth, we rephrased the results on locally annotated databases (Section~\ref{sec:locally-annotated}) to state the precise tractability condition rather than a procedure for its detection~\cite[Theorem 20]{DBLP:conf/icdt/EldarCK24}. In particular, this rephrasing allowed us to add \Cref{cor:min-max-tractability} for the min and max aggregations.

The remainder of the manuscript is organized as follows. After  preliminary concepts and notation in Section~\ref{sec:prelim},
Section~\ref{sec:direct} defines the challenge of direct access with an underlying order and recalls the state of affairs for ordinary CQs over ordinary databases. In Section~\ref{sec:outside}, we analyze queries where the aggregation or annotation does not participate in the lexicographic order. We study the incorporation of the aggregation and annotation in the order in Section~\ref{sec:inside}, and in Section~\ref{sec:locally-annotated} we study the implication of local annotation (with a general addition and an idempotent addition) on the complexity of direct access. We conclude in Section~\ref{sec:conclusions}.

\section{Preliminaries}
\label{sec:prelim}

We begin with preliminary notation and terminology that we use throughout the paper.

\paragraph{Databases and conjunctive queries.}

A \emph{schema} $\scs$ is a finite set $\set{R_1,\dots,R_k}$ of relation symbols. Each relation symbol $R$ is associated with an arity $\ar(R)$, which is a natural number. 
We assume a countably infinite set $\vals$ of \e{values} that appear in databases.
A \e{database} $D$ over a schema $\scs$ maps every relation symbol $R$ of $\scs$ to a finite relation $R^D \subseteq \vals^{\ar(R)}$. If $(c_1,\dots,c_k)$ is a tuple of $R^D$ (where $k=\ar(R)$), then we call the expression $R(c_1,\dots,c_k)$ a \e{fact} of $D$. 

A \emph{Conjunctive Query (CQ)} over the schema $\scs$ has the form 
$Q(\vec{x}) \datarule \varphi_1(\vec{x},\vec{y}),\dots,\varphi_\ell(\vec{x},\vec{y})$
where $\vec x$ and $\vec y$ are disjoint sequences of variables, and each 
$\varphi_i(\vec{x},\vec{y})$ is an \e{atomic query}
$R(z_1,\dots,z_k)$ such that $R\in\scs$ with $\ar(R)=k$ and each $z_i$ is  a variable in $\vec x$ or $\vec y$. Each
$\varphi_i(\vec x,\vec y)$ is an \e{atom} of $Q$, and we denote by $\atoms(Q)$ the set of atoms of $Q$.  We call $Q(\vec{x})$ the \e{head} of the query and $\varphi_1(\vec{x},\vec{y}),\dots,\varphi_\ell(\vec{x},\vec{y})$ the \e{body} of the query. The variables of $\vec x$ are the \e{free} variables of $Q$, and those of $\vec y$ are the \e{existential} variables of $Q$, and every variable occurs at least once in the body. We use $\var(Q)$ and $\free(Q)$ to denote the set of all variables and all free variables of $Q$, respectively. If $\varphi\in\atoms(Q)$, then $\var(\varphi)$ is the set of variables in $\varphi$. 
We say that $Q$ is \e{full} if it has no existential variables, that is $\var(Q)=\free(Q)$.

We refer to a database $D$ over the schema $\scs$ of the CQ $Q$ as a \e{database over $Q$}. A \e{homomorphism} from a CQ $Q$ to a database $D$ over $Q$ is a mapping $h$ from the variables of $Q$ into values of $D$ such that for each atom $R(z_1,\dots,z_k)$ of $Q$ it holds that
    $R(h(z_1),\dots,h(z_k))$ is a fact of $D$.
We denote by $\hom(Q,D)$ the set of all homomorphisms from $Q$ to $D$. 
If $h\in\hom(Q,D)$ then we denote by $h(\vec x)$ the tuple obtained by replacing every variable $x$ with the value $h(x)$, and we denote by $h(\varphi_i(\vec x,\vec y))$ the fact that is obtained from the atom $\varphi_i(\vec x,\vec y)$ by replacing every variable $z$ with the value $h(z)$. 
An \e{answer} to $Q$ over $D$ is a tuple of the form $h(\vec x)$ where $h\in\hom(Q,D)$. The \e{result} of $Q$ over $D$, denoted $Q(D)$, is 
$Q(D)\defeq \set{h(\vec x)\mid h\in\hom(Q,D)}$.

In our proofs, we will use the following definition of when two facts $f$ and $f'$ agree in the context of two queries. Intuitively, facts agree if they assign the same variables with the same values.
Let $Q$ and $Q'$ be CQs over the schemas $\scs$ and $\scs'$, respectively. Let $\varphi$ and $\varphi'$ be atoms of $Q$ and $Q'$ over the relation symbols $R$ and $R'$, respectively. Let $f$ and $f'$ be facts over $R$ and $R'$, respectively. 
We say that $f$ and $f'$ \e{agree} (w.r.t.~$\varphi$ and $\varphi'$) if there exists a homomorphism $h:\var(\varphi) \cup \var(\varphi')\rightarrow\vals$ such that $f$ and $f'$ are obtained from $\varphi$ and $\varphi'$, respectively, by replacing each variable $x$ with the value $h(x)$.

Consider a CQ $Q(\vec{x}) \datarule \varphi_1(\vec{x},\vec{y}),\dots,\varphi_\ell(\vec{x},\vec{y})$. We may refer to $Q$ as $Q(\vec x)$ to specify the sequence of free variables in the head. In this work, the \e{order} of the sequence $\vec{x}$ has a crucial role, since it determines the desired order of answers. Specifically, we will assume that the desired order of answers is \e{lexicographic} in the left-to-right order of $\vec{x}$. For example, the CQ $Q(x_1,x_2)\datarule R(x_1,x_2),S(x_2,y)$ differs from the CQ $Q'(x_2,x_1)\datarule R(x_1,x_2),S(x_2,y)$ not only in the order of values within each answer tuple but also in the order over the answers. For $Q(x_1,x_2)$ we order the answers first by $x_1$ and then by $x_2$, and for $Q'(x_2,x_1)$ we order first by $x_2$ and then by $x_1$.
 
As usual, we associate a CQ $Q$ with the hypergraph $H(Q) = (V_Q, E_Q)$ where $V_Q = \var(Q)$ and $E_Q = \{ \var(\varphi) | \varphi \in \atoms(Q) \}$. We say that $Q$ is \e{acyclic} if $H(Q)$ is an acyclic ($\alpha$-acyclic) hypergraph. Recall that a hypergraph $H = (V, E)$ is acyclic if there is a tree $T = (E, E_T)$, called a \e{join tree} of $H$, with the running intersection property: for each vertex $v\in V$, the set of hyperedges that contain $v$ induces a connected subtree of $T$. If $H$ is acyclic and $S\subseteq V$, then we say that $H$ is \e{$S$-connex} if $H$ remains acyclic even if we add $S$ to the set of hyperedges~\cite{braultbaron:tel-01081392}.
An acyclic CQ $Q$ is \e{free-connex} if $H(Q)$ is acyclic and $\free(Q)$-connex. 

A hypergraph $H' = (V,E')$ is an \e{inclusive extension} of a hypergraph $H=(V,E)$ if $E \subseteq E'$ and for  every edge $e'\in E'$ there is an edge $e\in E$ such that $e'\subseteq e$. It is known that $H$ is acyclic $S$-connex if and only if $H$ has an inclusive extension with a join tree $T$ such that $S$ is precisely the set of all variables contained in the vertices of some subtree of $T$~\cite{10.1007/978-3-540-74915-8_18}. We call such a tree ext-$S$-connex tree. 
When $S$ is the set of free variables of the CQ, and the CQ is clear from the context, we call such a tree \e{ext-free-connex}.

Let $Q$ be a CQ. Two variables of $Q$ are \e{neighbors}  (in $Q$) if they occur jointly in at least one of the atoms. If $\vec x$ is a set of variables in $\var(Q)$, then we denote by $\neigh_Q(\vec x)$ the \e{neighborhood} of $\vec x$ that consists of every variable in $\vec x$ and every neighbor of every variable in $\vec x$.  If $Q$ is clear from the context, then we may omit it and write just $\neigh(\vec x)$.

The notion of a \e{disruptive trio} has been introduced previously in the context of direct access to the answers of CQs~\cite{carmeli2023tractable}. A disruptive trio of a CQ $Q(\vec x)$ is a set of three distinct free variables $x_1$, $x_2$, and $x_3$ such that 
$x_1$ and $x_2$ neighbor $x_3$ but not each other, and $x_3$ succeeds both $x_1$ and $x_2$ in $\vec x$. 

\subparagraph*{Aggregate queries.}
By an \e{aggregate function} we refer to a function that takes as input a bag of tuples over $\vals$ and returns a single value in $\vals$. We adopt the notation of 
Cohen et al.~\cite{10.1145/1219092.1219093}, as follows. An \e{aggregate query} here is an expression of the form 
$$Q(\vec{x}, \alpha(\vec{w}),\vec{z}) \datarule  \varphi_1(\vec{x},\vec{y},\vec{z}),\dots,\varphi_\ell(\vec{x},\vec{y},\vec{z})$$
such that
$Q'(\vec{x},\vec{z}) \datarule  \varphi_1(\vec{x},\vec{y},\vec{z}),\dots,\varphi_\ell(\vec{x},\vec{y},\vec{z})$ is a CQ, $\alpha$ an aggregate function, and $\vec w$ a sequence of  variables from $\vec y$. An example is
$Q(x_1,x_2,\aggsum(y_2),z)\datarule R(x_1,x_2,y_1),S(y_1,y_2,z)$.
We refer to such a query as an \e{Aggregate CQ} or \e{\ACQ} for short. Given a database $D$ over $Q'$, the result $Q(D)$ is defined by
$Q(D)\defeq \set{(\vec{a},\alpha(B(\vec{a},\vec{b})),\vec{b})\mid (\vec{a},\vec{b})\in Q'(D)}$
where $B(\vec{a},\vec{b})$ is the bag that is obtained by collecting the tuples $h(\vec{w})$ from every $h\in\hom(Q',D)$ with
$h(\vec x)=\vec a$ and $h(\vec z)=\vec b$.  Note that our database and query model use set semantics, and we use bag semantics only to define the aggregate functions (in order to capture important functions such as count and sum).

We say that $Q$ is \e{acyclic} if $Q'$ is acyclic. Similarly, $Q$ is \e{free-connex} if $Q'$ is free-connex. A \e{disruptive trio} of $Q$ is a disruptive trio of $Q'$; in other words, the definition of a disruptive trio remains unchanged when introducing aggregates, while we consider only the free variables and not the aggregate function. 

\begin{rem}
We remark on two aspects in our definition of \ACQs. First, the reason for using both $\vec{x}$ and $\vec{z}$ as sequences of free variables is to determine a position for the aggregate value $\alpha(\vec{w})$ and, consequently, define its position in the lexicographic order over the answers. Second, the reader should note that, in our notation, an \ACQ has a single aggregate function. While this is important for some of our results, other results can be easily extended to multiple aggregate functions $\alpha(\vec{w_1})$, \dots, $\alpha(\vec{w_k})$. We will mention this extension when relevant.\qed
\end{rem}

In this work, we restrict the discussion to the common aggregate functions $\aggcount$, $\aggcountd$ (count distinct), $\aggsum$, $\aggavg$ (average), $\aggmin$, and $\aggmax$. All aggregate functions take a single column as input (i.e., $\vec{w}$ is of length one) except for 
$\aggcount$ that counts the tuples in the group and takes no argument. For instance, the query $Q_2$ in the Introduction uses $\aggcount()$ and it could also use
$\aggcountd(g)$ for counting the distinct games with scored goals.

\subparagraph*{Commutative semirings.}

A \e{commutative monoid} is an algebraic structure $(\K,\cdot)$ over a domain $\K$, with a binary operation $\cdot$ that satisfies 
 \e{associativity}: $(a \cdot b) \cdot c = a \cdot (b \cdot c)$ for any $a,b,c \in \K$, \e{commutativity}: $a \cdot b = b \cdot a$ for any $a,b \in \K$, and \e{identity element}: there exists an element  $\varnothing \in \K$ such that $a \cdot \varnothing = a$ for any $a \in \K$.
A \e{commutative semiring} is an algebraic structure $(\K,\oplus,\otimes,\zero,\one)$ over a domain $\K$, with two binary operations $\oplus$  and $\otimes$  and two distinguished elements $\zero$  and $\one$ in $\K$ that satisfy the following conditions:
  \e{(a)} $(\K, \oplus)$ is a commutative monoid with the identity element $\zero$;
  \e{(b)} $(\K, \otimes)$ is a commutative monoid with the identity element $\one$; 
  \e{(c)} $a \otimes (b \oplus c) = (a \otimes b) \oplus (a \otimes c)$ for all $a,b,c \in\K$; 
  \e{and (d)} $a \otimes \zero = \zero$ for all $a \in \K$.

We refer to $\oplus$ as the \e{addition} operation, $\otimes$ as the \e{multiplication} operation, $\zero$ as the \e{additive identity} and $\one$ as the  \e{multiplicative identity}. We give examples of commutative semirings at the end of this section.

\subparagraph*{Annotated databases and query answers.}

Let $\scs$ be a schema and $(\K,\oplus,\otimes,\zero,\one)$ a commutative semiring. A \e{$\K$-database} (\e{over $\scs$}) is a pair $(D,\tau)$ where $D$ is a database over $\scs$ and $\tau:D\rightarrow\K$ is function that maps every fact $f$ of $D$ to an element $\tau(f)$ of $\K$, called the \e{annotation} of $f$. 

The annotation of a database propagates to the query answers by associating a semiring operation with each algebraic operation~\cite{10.1145/1265530.1265535}. In the case of CQs, the relevant operations are \e{joins} and \e{projection}. For join,  the annotation of the result is the product of the annotation of the operands. For projection, the annotation is the sum of the annotations of the tuples that give rise to the answer. In our terminology, we have the following. 

Let $Q(\vec{x}) \datarule \varphi_1(\vec{x},\vec{y}),\dots,\varphi_\ell(\vec{x},\vec{y})$ 
be a CQ and $(D,\tau)$ an annotated database. 
For a homomorphism $h$ from $Q$ to $D$, we denote by $\otimes h$ the product of the annotations of the facts in the range of $h$, that is
$\otimes h\defeq \tau(h(\varphi_1(\vec{x},\vec{y})))\otimes\dots\otimes \tau(h(\varphi_\ell(\vec{x},\vec{y})))$.
An \e{answer} to $Q$ over $(D,\tau)$ is a pair $(\vec c,a)$ such that $\vec c\in Q(D)$ and
    $$a=\oplus\set{\otimes h\mid h\in\hom(Q,D) \land h(\vec x)=\vec c}$$
where, for $A=\set{a_1,\dots,a_n}\subseteq\K$, we define 
$\oplus A=a_1\oplus\dots\oplus a_n$.
As before, the \e{result} of $Q$ over $(D,\tau)$, denoted 
$Q(D,\tau)$, is the set of answers $(\vec c,a)$ to $Q$ over $(D,\tau)$.
We will make use of the fact that, over commutative semirings, projections and joins are commutative~\cite{10.1145/1265530.1265535}.

In this work, we study the ability to incorporate the annotation in the order over the answers. More precisely, we will investigate the complexity of involving the annotation in the lexicographic order over the answers, as if it were another value in the tuple. So, when we consider a CQ $Q(\vec{x})$, we need to specify where the annotation goes inside $\vec{x}$. Similarly to the \ACQ notation, we do so by replacing $\vec{x}$ with a sequence $(\vec x,\star,\vec z)$ where $\star$ represents the annotation value. We refer to a CQ of this form as a \cqstar. An example of a \cqstar is
$Q(x_1,x_2,\star,z)\datarule R(x_1,x_2,y_1),S(y_1,y_2,z)$
where the lexicographic order is by $x_1$, then by $x_2$, then by the annotation, and then by $z$.

\begin{rem}
The literature on the complexity of query evaluation over annotated databases sometimes considers answers  $(\vec c,a)$ as valid only if $a \neq \zero$. For instance, this convention is adopted in the recent work of Mu{\~{n}}oz, Riveros, and Vansummeren~\cite{DBLP:journals/lmcs/MunozRV26} on conjunctive queries for linear algebra over matrices over a semiring. In contrast, our semantics allows answers with an annotation of zero. This variation is significant in our context, since the number of answers with a zero annotation can be large. The techniques developed in this manuscript do not readily extend to supporting this restriction, and we leave this direction for future work.
\qed
\end{rem}

Let $Q$ be a \cqstar, and let $Q'$ be the CQ obtained from $Q$ by removing $\star$ from the head. As in the case of \ACQs, $Q$ is \e{acyclic} if $Q'$ is acyclic, $Q$ is \e{free-connex}  if $Q'$ is free-connex, and a \e{disruptive trio} of $Q$ is a disruptive trio of $Q'$.

Aggregate functions can often be captured by annotations of answers in annotated databases, where each aggregate function might require a different commutative semiring:
\begin{itemize}
  \item $\aggsum$: the numeric semiring $(\Q,+,\cdot,0,1)$.
  \item $\aggcount$: the counting semiring $(\N,+,\cdot,0,1)$.
  \item $\aggmax$: the max tropical semiring $(\Q \cup \{ -\infty \},\maxf, +, -\infty, 0)$.
  \item $\aggmin$: the min tropical semiring $(\Q \cup \{ \infty \},\minf, +, \infty, 0)$.
\end{itemize}

The translation is straightforward (and known, e.g.,~\cite{DBLP:journals/vldb/ReS09,DBLP:journals/tods/KhamisCMNNOS20}), as we illustrate in \Cref{fig:aggregate-to-annotation}: the aggregated value becomes the annotation on one of the relations, the annotation outside of this relation is the multiplicative identity (as we later term ``locally annotated''), and the addition operation captures the aggregate function. Note that in the case of the numeric and min/max tropical semirings, we are using the domain $\Q$ of rational numbers rather than all real numbers to avoid issues of numerical presentation in the computational model.

In general, the translation is directly applicable whenever the aggregate function is defined as a commutative monoid over the elements of a column. It may also be applicable indirectly. For example,
$\aggavg$ can be computed using $\aggsum$ and $\aggcount$. $\aggcountd$ (count distinct) cannot be captured by a semiring, as the result of $\oplus$ cannot be computed from two intermediary annotations in the domain. We can, however, capture a semantically similar concept with the set semiring $(\powerset(\cntdomain),\cup, \cap, \varnothing, \cntdomain)$ by annotating each fact with the actual set of distinct elements. However, in such cases, we will need our complexity analysis to be aware of the cost of the operations.

\begin{figure}
  \small
\begin{tabular}{cc}
  \multicolumn{2}{l}{\rel{Teams}}\\\toprule
  $p$ & $c$ \\\midrule
  1 & 5\\
  2 & 5\\
  3 & 6\\
  4 & 7\\
  5 & 8\\
  \bottomrule
\end{tabular}
\begin{tabular}{ccc}
  \multicolumn{3}{l}{\rel{Goals}}\\\toprule
   $g$ & $p$ & $t$ \\\midrule
  1 & 1 & 31\\
  1 & 3 & 50\\
  1 & 3 & 75\\
  2 & 4 & 90\\
  2 & 4 & 9\\
  \bottomrule
\end{tabular}\quad
\begin{tabular}{cc}
  \multicolumn{2}{l}{\rel{Replays}}\\\toprule
   $g$ & $t$ \\\midrule
  1 & 1\\
  1 & 31\\
  1 & 50\\
  2 & 5\\
  1 & 90\\
  \bottomrule
\end{tabular}
\quad$ \Rightarrow$
\quad
\begin{tabular}{cc>{\columncolor{gray!30}}c}
  \multicolumn{3}{l}{\rel{Team}}\\\toprule
  $p$ & $c$  & $\tau_+$\\\midrule
  1 & 5 & 1\\
  2 & 5 & 1\\
  3 & 6 & 1\\
  4 & 7 & 1\\
  5 & 8 & 1\\
  \bottomrule
\end{tabular}\quad
\begin{tabular}{ccc>{\columncolor{gray!30}}c}
  \multicolumn{4}{l}{\rel{Goals}}\\\toprule
   $g$ & $p$ & $t$ & $\tau_{+}$\\\midrule
  1 & 1 & 31 & 1\\
  1 & 3 & 50 & 1\\
  1 & 3 & 75 & 1\\
  2 & 4 & 90 & 1\\
  2 & 4 & 9 & 1\\
  \bottomrule
\end{tabular}\quad
\begin{tabular}{cc>{\columncolor{gray!30}}c}
  \multicolumn{3}{l}{\rel{Replays}}\\\toprule
   $g$ & $t$ & $\tau_{+}$\\\midrule
  1 & 1 & 1\\
  1 & 31 & 31\\
  1 & 50 & 50\\
  2 & 5 & 5\\
  1 & 90 & 90\\
  \bottomrule
\end{tabular}
 \caption{An example of a $\Q$-database over the numerical semiring constructed to evaluate the \ACQ $Q(c, \aggsum(t)) \datarule \rel{Teams}(p,c), \rel{Goals}(g,p,t),\rel{Replays}(g,t)$. }\label{fig:aggregate-to-annotation}
\end{figure}

\section{The Direct-Access Problem}\label{sec:direct}
In this paper, we study CQs with lexicographic orders over the answers. As said earlier, the lexicographic order for the CQ $Q(\vec x)$ is left to right according to $\vec x$. We will also investigate lexicographic orders that involve the annotation or aggregation when the query is a \cqstar 
$Q(\vec x,\star,\vec z)$ or an \ACQ $Q(\vec x,\alpha(\vec w),\vec z)$, respectively. 
We refer uniformly to the annotation of an answer (over an annotated database) and to the aggregate value of the answer's group (over an ordinary database) as the \e{computed value}. 

Let $Q$ be a CQ, \cqstar, or an \ACQ. A \e{direct access} solution for $Q$ consists of two algorithms: one for \e{preprocessing} and one for \e{access}. 
\begin{itemize}
\item The preprocessing algorithm takes as input a database $D$ over $Q$ and constructs a data structure $S_D$.
\item The access algorithm takes as input $S_D$ and an index $i$, and returns the $i$th answer of $Q(D)$ in the lexicographic order. Note that this answer includes the computed value, when it exists. If $i>|Q(D)|$ then the algorithm should return \e{null}.
\end{itemize}

To define the complexity requirements of \e{efficient} direct access, we first describe the complexity model that we adopt. We use \emph{data complexity} as a yardstick of tractability. Hence, complexity is measured in terms of the size of the database, while the size of the query is fixed (and every query is a separate computational problem). Assuming the input is of size $n$, we use the RAM model of computation with $O(\log n)$-bit words and uniform-cost operations. Notably, this model allows us to assume perfect hash tables can be constructed in linear time, and they provide access in constant time~\cite{louis-model}. 

Regarding the complexity of the semiring operations, the RAM model allows us to assume that the numeric, counting, min tropical, and max tropical semirings use constant space for representing values and constant time for the operations $\oplus$ and $\otimes$. In fact, it suffices for our results to assume that the operations take logarithmic time, and later we will make use of this relaxed assumption (within a special case of $\aggcountd$). We refer to a semiring with this property as a \e{logarithmic-time (commutative) semiring}.

Let $T_p$ and $T_a$ be numeric functions.
A direct-access algorithm is said to be \e{in $\angs{T_p,T_a}$} if, for every input database $D$, the preprocessing phase takes $O(T_p(|D|))$ time and each access takes $O(T_a(|D|))$ time. For example, $\linlog$ states that preprocessing constructs in $O(|D|\log|D|)$ time a data structure that provides $O(\log|D|)$-time access.
In this work, a query $Q$ has \e{efficient} direct access (and $Q$ is deemed tractable) if it has a direct access algorithm in $\linlog$.

Carmeli et al.~\cite{carmeli2023tractable} established a dichotomy in the tractability of the CQs and lexicographic orders.
This dichotomy relies on the following hypotheses.
\begin{itemize}
\item \sparseBMM: two binary matrices $A^{n \times n}$ and $B^{n \times n}$ represented by lists of their non-zero entries cannot be multiplied in $O(m \polylog(m))$ time where $m$ is the total number of non-zero entries in $A$, $B$ and $A \times B$. 
\item \HYPERCLIQUE: for all $k\ge 2$, there is no $O(m \polylog(m))$-time algorithm that, given a hypergraph with $m$ hyperedges, determines whether there exists a set of $k+1$ vertices such that every subset of $k$ vertices among them forms a hyperedge.
\end{itemize}

\begin{thmC}[\cite{carmeli2023tractable}]\label{thm:known-dichotomy}
Let $Q$ be a CQ.
\begin{enumerate}
    \item If $Q$ is free-connex with no disruptive trio, then direct access for $Q$ is in $\linlog$.
    \item Otherwise, if $Q$ is also self-join-free, then direct access for $Q$ is not in $\linlog$, assuming the
    \HYPERCLIQUE hypothesis (in case $Q$ is cyclic) and the \sparseBMM hypothesis (in case $Q$ is acyclic).
\end{enumerate}
\end{thmC}

The following lemma claims that introducing new variables into a hard query cannot make it easier (as long as the connections between the original variables remain unchanged).

To formalize this claim, we use the following definitions.
Given a self-join-free query $Q$ and a subset $V$ of its variables, we denote by $Q_{|V}$ the query obtained by removing all variables that are not in $V$ from $Q$ (thus, the arity of the atoms or head may decrease).
We say that two queries $Q_1$ and $Q_2$ are \emph{structurally equivalent} if there exists a query $Q_1'$ that is isomorphic to $Q_1$ (i.e., $Q_1'$ is obtained by renaming the variables of $Q_1$) such that: (1) the variables of every atom of $Q_1'$ are contained in an atom of $Q_2$ and vice versa; and (2) $Q_1'$ and $Q_2$ have the same head.

\begin{lem}\label{lem:reduction}
    Let $Q$ and $Q'$ be queries such that they are both CQs, both \cqstars over a logarithmic-time semiring, or both \ACQs, and $Q$ has no self-joins. 
    If there exists $V\subseteq\var{(Q)}$ such that $Q_{|V}$ and $Q'$ are structurally equivalent, and direct access for $Q$ is in $\angs{T_p,T_a}$, then direct access for $Q'$ is in $\angs{\loglin+T_p,T_a}$.
\end{lem}

As an example for the use of \Cref{lem:reduction}, it can reprove the fact that, given that $Q_\text{trio}(x_1,x_2,x_3)\datarule R_1(x_1,x_3),R_2(x_2,x_3)$ does not admit efficient direct access (assuming the \sparseBMM hypothesis), neither does any other CQ with a disruptive trio. We will later use \Cref{lem:reduction} similarly with different hard queries instead of $Q_\text{trio}$.

\begin{proof}[Proof of \Cref{lem:reduction}]
For the simplicity of presentation, we assume that the variables of $Q$ are renamed such that the variables of every atom of $Q_{|\var{(Q')}}$ are contained in an atom of $Q'$ and vice versa, and $Q_{|\var{(Q')}}$ and $Q'$ have the same head.

We show an \emph{order-preserving exact reduction} from $Q'$ to $Q$. That is, given a database $D'$ for $Q'$ we show how to construct a database $D$ for $Q$ in loglinear time such that there is an order-preserving bijection from $Q(D)$ to $Q'(D')$ and, moreover, the translation of a tuple from $Q(D)$ to its corresponding tuple from  $Q'(D')$  can be done in constant time. This proves that direct access for $Q$ implies direct access for $Q'$ with the same time guarantees after loglinear preprocessing.

For every atom $R(\vec{u})$ of $Q$, take an atom $R'(\vec{v})$ of $Q'$ that contains $\vec{u}$ restricted to $\var{(Q')}$. Construct a relation for $R$ from $R'$ according to these atoms while padding facts with the value $c$ where a value is missing. 
That is, treat every fact $R'(\vec{a})$ as a function $f$ mapping $v_i$ to $a_i$, extend this function to map all other variables to $c$, and add the fact $R(f(\vec{u}))$ to the construction.

Note that, for an atom $R(\vec{u})$ of $Q$, multiple atoms $R'(\vec{v})$ of $Q'$ may contain $\vec{u}$ restricted to $\var{(Q')}$. Since we select only one $R'(\vec{v})$, there may be some atoms of $Q'$ that do not participate in the above process.
For every atom $R'(\vec{v})$ of $Q'$ that was not used in the process, take an atom $R(\vec{u})$ of $Q$ that contains its variables, and filter its relation in $D$ according to $R'$. That is, for every fact $R(\vec{a})$ in our constructed instance, treat it as a function $f$ mapping $v_i$ to $a_i$, and check whether there is a corresponding fact $R'(f(\vec{u}))$ in the input; if none exists, then remove the fact $R(\vec{a})$ from the construction.

In case the queries are \cqstars, the database $D'$ is accompanied by the annotations $\tau'$, and we construct the annotations $\tau$ for $D$ as follows:
\begin{enumerate}
\item We first annotate all facts in the constructed instance with $\one$. 
\item Then, for every atom $R'(\vec{v})$ in $Q'$, we take an atom $R(\vec{u})$ in $Q$ containing its variables, and multiply the annotation of every fact in $R$ with the annotation of the corresponding fact in $R'$ (if a corresponding fact exists; if none exist, then this fact of $R'$ would not be used in an answer, and the annotation we choose for this fact does not matter).
\end{enumerate}
Note that the construction uses a linear number of operations. As the semiring multiplication takes logarithmic time, and the other operations take constant time, the construction takes $O(|D'|\log|D'|)$ time in total.

We claim that there is a bijection between homomorphisms from $Q'$ to $D'$ and homomorphisms from $Q$ to $D$ given by extending the homomorphism by mapping all other variables to $c$. In addition, in case the queries are \cqstars, multiplying all annotations of facts matching the source homomorphism and doing the same for the target homomorphism gives the same result.
In case the queries are \ACQs, this implies that the aggregation is applied to the same set of assignments, and therefore the aggregates computed in both cases are the same.
Together with the fact that $Q_{|\var{(Q')}}$ and $Q'$ have the same heads, this means that projecting $Q(D)$ to the variables of $Q'$ results in $Q'(D')$ with the same answer order (and the same annotations as in case of \cqstars), and this projection can be done in constant time.
\end{proof}

Next, we use \Cref{lem:reduction} to eliminate the existential variables from our queries, by setting $V=\free{(Q)}$.
It is folklore that free-connex CQs (over non-annotated databases) can be transformed into \e{full} acyclic CQs in linear time~\cite{10.1145/3035918.3064027, 10.1145/2656335}.
The following lemma states that the same holds for free-connex \cqstars over annotated databases.

\begin{lem}\label{lemma:existential-elimination-semiring-Qfree}
Let $(\K,\oplus,\otimes,\zero,\one)$ be a logarithmic-time commutative semiring, and let $Q(\vec x,\star,\vec z)$ be a self-join-free free-connex \cqstar. Then, $\Qpfree$ is a full acyclic \cqstar, and there is an $O(|D|\log|D|)$-time construction that maps $\K$-databases $(D, \tau)$ of $Q$ to $\K$-databases $(D', \tau')$ of $\Qpfree$ such that $\Qpfree(D', \tau') = Q(D, \tau)$.
\end{lem}

\begin{proof}
We show an $O(|D|\log|D|)$-time construction that maps $\K$-databases $(D, \tau)$ of $Q$ to $\K$-databases $(D', \tau')$ of $\Qpfree$ such that $\Qpfree(D', \tau') = Q(D, \tau)$.

Since $Q$ is free-connex, it has an ext-free-connex join tree $T$ with a subtree $T'$ which contains precisely the free variables of $Q$. We call the vertices of $T'$ the \emph{free vertices}.
The proof idea is as follows: we first adapt the database to have a relation for each vertex of $T$, we then build a database with the same answers using only the vertices of $T'$, and finally we translate this into a database for $\Qpfree$.

The first step is to make it so that each vertex of the tree has a unique associated atom in the query and a unique associated relation in the database.
For every vertex of the tree that does not correspond to a relation, we take the relation of a vertex that contains it (without the annotations), project it accordingly, and annotate the facts with $\one$.
The query answers are unchanged by the addition of these annotated relations as they contain projections of existing relations annotated with the multiplicative identity.

Next, our goal is to only use relations matching the free vertices and maintain the same query answers.
We claim that as long as $T$ has vertices that are not in the set of free vertices, we can remove such a vertex from $T$ and remove its associated atom from the query while adapting the database so that the query results remain unchanged.

As long as $T$ has a vertex not in the set of free vertices, eliminate such a leaf vertex as follows:
\begin{enumerate}
    \item In the relation associated with the vertex, project out every variable not shared between the vertex and its neighbor while summing ($\oplus$) annotations.\label{line:leaf-elimination-project}
    \item Join the obtained relation into the relation of the neighbor vertex while multiplying ($\otimes$) annotations.\label{line:leaf-elimination-join}
    \item Remove the leaf from the tree, its associated atom from the query, and its associated relation from the database.\label{line:leaf-elimination-node-removal}
\end{enumerate}

More specifically,
denote by $S$ the variables shared between the treated leaf vertex $v$ and its neighbor. 
During step~(\ref{line:leaf-elimination-project}), we initialize a lookup table where for each unique assignment of $S$ matching a fact in the relation of $v$, we set the semiring value $\zero$.
Then, we iterate over each fact of the relation of $v$ and add its annotation (using the semiring $\oplus$) to the entry matching its assignment to $S$.
Next, during step~(\ref{line:leaf-elimination-join}), we iterate over every fact in the relation matching the neighbor of $v$ and multiply its existing annotation (using the semiring $\otimes$) with the value stored for its assignment to $S$.
Once we are done, each fact in the relation matching the neighbor of $v$ had its annotation multiplied by the sum of the annotations of the facts that agree with it in the relation of $v$.
In terms of complexity, we perform the elimination of a single vertex in loglinear time. Indeed, the computation model allows constructing in linear time lookup tables that can be accessed in constant time, and we perform a linear number of semiring operations, each taking logarithmic time. Since the number of elimination steps is bounded by the size of the query, the entire elimination takes loglinear time.

As for the correctness, consider the variables that appear in $v$ but not in its neighbor. 
Since $v$ is chosen to be a leaf, the running intersection property guarantees that these variables only appear in $v$. This means that they do not appear in the free vertices, so they are existential and must be projected out. As joins and projections are commutative over annotated databases \cite{10.1145/1265530.1265535}, we can do the projection first. As these variables appear only in $v$, the projection can be done locally. Thus, the results of the modified query over the modified database remain unchanged.

Finally, we adapt the relations matching the free vertices into a database for $\Qpfree$.
We first claim that for each atom $\varphi$ of $Q$, we get that $\free(\varphi)$ is contained in some free vertex. As $T$ is an ext-free-connex tree, $\var(\varphi)$ is a vertex of $T$. If $\var(\varphi)$ is a free vertex, then $\var(\varphi) = \free(\varphi)$ is a free vertex. Otherwise, consider the nearest free vertex to $\var(\varphi)$. Since $T'$ contains all free variables of $Q$, in particular $\free(\varphi)$, then from the running intersection property, we know that the nearest vertex must also contain $\free(\varphi)$.

Consider a vertex of $T'$ which does not match an atom of $\Qpfree$. Since it is a vertex of $T'$, we know that it only contains free variables, and that its variables are contained in some atom of $Q$. This means that its variables are also contained in some atom of $\Qpfree$ and thus some other vertex of $T'$. Therefore, we can eliminate $v$ (and its relation counterpart) by joining its relation into the relation of the containing vertex in the same fashion as in step~(\ref{line:leaf-elimination-join}).
Next, consider an atom of $\Qpfree$ which does not have a matching vertex in $T'$. We know that $T'$ has a vertex $u$ containing the variables of this atom. For this atom, we take the relation of $u$ (without the annotations), project it accordingly, and annotate the facts with $\one$.
As we only added facts projected from existing relations and annotated by $\one$, the query answers and annotations remain unchanged.
\end{proof}

We note that in the case where the semiring operations require only constant time, or in the case that \Cref{lem:reduction} is applied to CQs or \ACQs, the construction time of \Cref{lem:reduction} and \Cref{lemma:existential-elimination-semiring-Qfree} is linear time instead of loglinear time. 
In addition, when using \Cref{lemma:existential-elimination-semiring-Qfree}, two free variables are neighbors in $\Qpfree$ if and only if they are neighbors in $Q$, and since the order remains the same, $\Qpfree$ will have a disruptive trio if and only if $Q$ has a disruptive trio.

\Cref{lemma:existential-elimination-semiring-Qfree} shows that, if direct access for $\Qpfree$ over $\K$-databases is in $\angs{T_p,T_a}$, then direct access for $Q$ over $\K$-databases is in $\angs{\loglin+T_p,T_a}$.
By combining this with \Cref{lem:reduction} (for $V=\free{(Q)}$), we get that $Q$ and $\Qpfree$ are equally hard for direct access.
\begin{thm}\label{thm:existential-removal}
    Let $(\K,\oplus,\otimes,\zero,\one)$ be a logarithmic-time commutative semiring, and let $Q(\vec x,\star,\vec z)$ be a self-join-free free-connex \cqstar. Then, direct access for $Q$ is in $\angs{T_p,T_a}$ if and only if direct access for $\Qpfree$ is in $\angs{T_p,T_a}$, assuming $T_p=\Omega(\mathrm{loglinear})$.
\end{thm}

The positive side of \Cref{thm:existential-removal} can easily be used also in case $Q$ contains self-joins by removing the self-joins while duplicating the relevant relations.
We say that a query $\Qsf$ is a self-join-free version of a query $Q$ if it can be obtained from
$Q$ by replacing the relation symbols such that each relation symbol appears at most once in $\Qsf$.

\begin{cor}\label{cor:pos-existential-removal}
Let $(\K,\oplus,\otimes,\zero,\one)$ be a logarithmic-time commutative semiring, and let $Q(\vec x,\star,\vec z)$ be a free-connex \cqstar. Let $\Qsf$ be a self-join-free version of $Q$. If direct access for $\Qsffree$ is in $\linlog$, then direct access for $Q$ is in $\linlog$.
\end{cor}

\section{Incorporating Annotation and Aggregation in the Answers}\label{sec:outside}

In this section, we discuss the existence of efficient direct access in the case where the order \e{does not} involve the computed value, that is, the annotation (for \cqstars) or the aggregate values (for \ACQs). 
Equivalently, these are queries where the computed value is last in order, that is, the vector $\vec{z}$ in the head is empty. 
Hence, we focus on \cqstars of the form $Q(\vec x,\star)$ and \ACQs of the form $Q(\vec x,\alpha(\vec w))$. 
In other words, the problem is similar to the CQ case, except that the access algorithm should also retrieve the aggregated value from the data structure. In 
Section~\ref{sec:extending-dichotomy}, we will use annotated databases to identify the cases where this can be done efficiently for min, max, count, sum, and average. 
By contrast, in Section~\ref{sec:count-distinct} we will show that for count-distinct, the class of tractable cases is more restricted unless the domain of the elements we count is small (logarithmic size).

\subsection{Generalized Dichotomies}\label{sec:extending-dichotomy}

We now show that \Cref{thm:known-dichotomy} extends to databases with annotations, and so, also to queries with aggregate functions that can be efficiently simulated by annotations. 

\begin{thm}\label{thm:annotated-dichotomy}
Let $(\K,\oplus,\otimes,\zero,\one)$ be a logarithmic-time commutative semiring, and let $Q(\vec x,\star)$ be a \cqstar.
\begin{enumerate}
    \item If $Q$ is free-connex with no disruptive trio, then direct access for $Q$ is in $\linlog$ on $\K$-databases.
    \item Otherwise, if $Q$ is also self-join-free, then direct access for $Q$ is not in $\linlog$, assuming the
    \HYPERCLIQUE and \sparseBMM hypotheses.
\end{enumerate}
\end{thm}

\begin{proof}

For the negative side of the dichotomy, we simply use the negative side of \Cref{thm:known-dichotomy}. This can be done since each answer to a \cqstar contains the ordinary (non-annotated) answer to the CQ obtained by removing $\star$, and the answers have the same order.

For the positive side of the dichotomy,  \Cref{cor:pos-existential-removal} implies that it is enough to show an algorithm for $Q'(\vec x,\star)=\Qsffree$, which is a full self-join-free acyclic \cqstar with no disruptive trios.
Using \Cref{thm:known-dichotomy}, we have direct access for $Q'$ in $\linlog$ if we ignore the annotations. Since $Q'$ is full, from an answer to $Q'$ we compute the annotation by finding the tuples that agree with the answer in every relation of $Q'$ and multiply the annotations of these tuples; this can be done in logarithmic time.
\end{proof}

From the positive side of \Cref{thm:annotated-dichotomy}, we conclude efficient direct access for \ACQs $Q(\vec{x}, \alpha(\vec w))$,
as long as we can efficiently formulate the aggregate function as an annotation over some logarithmic-time commutative semiring. This is stated in the following corollary of \Cref{thm:annotated-dichotomy}.

\begin{cor}\label{cor:aggregate-tractable}
Consider an \ACQ
$Q(\vec{x}, \alpha(\vec w)) \datarule  \varphi_1(\vec{x},\vec{y}),\dots,\varphi_\ell(\vec{x},\vec{y})$
where $\alpha$ is one of $\aggmin$, $\aggmax$, $\aggcount$, $\aggsum$, and $\aggavg$.
\begin{enumerate}
    \item If the CQ $Q'(\vec{x}) \datarule \varphi_1(\vec{x},\vec{y}),\dots,\varphi_\ell(\vec{x},\vec{y})$ is free-connex with no disruptive trio, then direct access for $Q$ is in $\linlog$.
    \item Otherwise, if $Q$ is also self-join-free, then direct access for $Q$ is not in $\linlog$, assuming the
    \HYPERCLIQUE and \sparseBMM hypotheses.
\end{enumerate}
\end{cor}

\begin{proof}
For the positive side, we simply apply \Cref{thm:annotated-dichotomy} with the corresponding semiring. 
In the case where $\alpha$ is $\aggavg$, we compute $\aggsum$ and $\aggcount$ separately and divide the results.
The negative side carries over from \Cref{thm:known-dichotomy} since a direct access solution for $Q$ 
in $\linlog$ is also a direct access solution for $Q'$ in $\linlog$ if we ignore the aggregated values.
\end{proof}

\begin{rem}
\Cref{cor:aggregate-tractable} can be easily extended to support multiple aggregate functions
$\alpha_1(\vec w_1)$, \dots, $\alpha_k(\vec w_k)$.  For that, we can simply solve the problem for each 
$\alpha_i(\vec w_i)$ separately, and extract the aggregate values of an answer from the $k$ data structures that we construct in the preprocessing phase. (Moreover, a practical implementation can handle all aggregate values in the same structure.)
\qed
\end{rem}

\subsection{Count Distinct}\label{sec:count-distinct}

Can we generalize \Cref{cor:aggregate-tractable} beyond the stated aggregate functions? The most notable missing aggregate function is $\aggcountd$ (count-distinct).
Next, we show that we \e{cannot} have similar tractability for count-distinct, and we illustrate it with a specific query (\Cref{lem:hsc}). 
After that, we show how precisely \Cref{cor:aggregate-tractable} changes for count-distinct, and particularly which \ACQs move from the tractable to the intractable side (\Cref{thm:countd-dichotomy}). We later explain that the tractable cases of \Cref{cor:aggregate-order-tractable} are fully restored if we assume that the domain size is logarithmic in that of the input (\Cref{rem:countd-small-domain}).

The negative results in this section require the \emph{small-universe Hitting Set Conjecture (\HSC)}~\cite{vassilevska2015hardness}. In \HSC, we are given two sets $\U$ and $\V$ of size $N$, each containing sets over the universe $\{ 1, 2, \ldots, d \}$, and the goal is to determine whether $\U$ contains a set that hits (i.e., shares an element with) every set in $\V$. \HSC states that the problem takes $N^{2-o(1)}$ time for every function $d=\omega(log(N))$. (In fact, it is conjectured that even a randomized algorithm for this problem needs $N^{2-o(1)}$ time in expectation~\cite{vassilevska2015hardness}.)

\begin{lem}\label{lem:hsc}
Assuming \HSC, for all $k\geq 1$, direct access is not in $\linlog$ for $Q_H^k(x, \aggcountd(w)) \datarule R_0(x,y_1), R_1(y_1,y_2),\ldots,R_{k-1}(y_{k-1},y_k), R_k(y_k,w)$. 
\end{lem}

\begin{proof}
Let $\U = \set{U_1,U_2,\ldots,U_N}$ and $\V=\set{V_1,V_2,\ldots,V_N}$ be sets of sets of elements of the universe $\set{1,2, \ldots, d}$ where $d=N^c$ for some $0<c<1$. 
Indeed, $d=\omega(log N)$, as required by the conjecture. We construct a database with the fact $R_0(i,j)$ for all $j \in U_i$, the fact $R_k(j,i)$ for all $j \in V_i$, and the facts $R_i(j,j)$ for all $j$ in the universe and $0<i<k$.

Next, we assume efficient direct access for $Q_H^k$, and use that to solve the hitting-set problem.
Each query answer is a pair $(i,n)$ where $i$ is the index of a set $U_i$ from $\U$ and $n$ is the number of sets $V_j$ that $U_i$ hits.
By accessing all query answers, we can check all sets in $\U$, one by one, and test whether any is hitting all $N$ sets.
The number of facts in $D$ is $O(dN+d^2)\leq O(dN)$,
and the number of query answers (and so the number of access calls) is $N$.
Hence, direct access for $Q_H^k$ in $\linlog$ would imply a solution to the hitting-set problem in better than $N^{2-o(1)}$ time. Indeed, for any $d=O(N^c)$, we get a solution in $O(N^{1+c} \cdot\log N)$ time, which contradicts \HSC for $c<1$.
\end{proof}
Importantly, the reduction used in the proof of \Cref{lem:hsc} does not involve the order, and hence, the theorem holds true even for direct access, or enumeration, without any order requirement.

Next, we generalize the hardness proof from \Cref{lem:hsc} to a general class of 
\ACQs with count-distinct, and we devise an algorithm for other count-distinct queries. We show that this results in a dichotomy for count-distinct. The dichotomy states the following.
For the aggregate functions of \Cref{cor:aggregate-tractable},  we have established that the same dichotomy for standard CQs applies when we ``ignore'' the aggregation in the head and treat it as an \e{existential} variable. The case of count-distinct turns out to be the opposite: the dichotomy for standard CQs applies if we treat the aggregated variable as a \e{free} variable.

\begin{thm}\label{thm:countd-dichotomy}
Consider an \ACQ
$Q(\vec{x}, \aggcountd(w)) \datarule  \varphi_1(\vec{x},\vec{y}),\dots,\varphi_\ell(\vec{x},\vec{y})$.
\begin{enumerate}
    \item If the CQ $Q'(\vec{x},w) \datarule \varphi_1(\vec{x},\vec{y}),\dots,\varphi_\ell(\vec{x},\vec{y})$ is free-connex with no disruptive trio, then direct access for $Q$ is in $\linlog$.
    \item Otherwise, if $Q$ is also self-join-free, then direct access for $Q$ is not in $\linlog$, assuming the
    \HYPERCLIQUE, \sparseBMM, and \HSC hypotheses.
\end{enumerate}
\end{thm}

\def\Qfull{Q^\text{full}}
\def\Qcount{Q^\text{count}}
\begin{proof}
We start with the positive side of the dichotomy, and assume we are given a database $D$ as input.
First, we remove self-joins and eliminate the existential variables that are not part of the aggregation.
We can remove self-joins by replacing the relation symbols with unique relation names and copying the relations into their new names.
Since $Q'(\vec x,w)$ is free-connex with no disruptive trios, we can build in $O(|D|\log|D|)$ time a full acyclic CQ $\Qfull(\vec x,w)$, without self-joins or disruptive trios, and a database $D'$ for $\Qfull$ such that $\Qfull(D') = Q'(D)$.
This transformation is folklore and can be seen as a special case of our construction from \Cref{lemma:existential-elimination-semiring-Qfree} when we ignore the annotation. We illustrate the construction in \Cref{example:theorem-countd} following the proof.

As $\Qfull$ has no disruptive trio, we know that it has an atom that contains its last free variable and all of its neighbors~\cite[proof of Lemma 3.9]{carmeli2023tractable}.
In other words, when considering all the atoms that contain $w$, there is a maximal atom $\varphi$ that contains all of their variables. We filter the relation of $\varphi$ according to all of these atoms. That is, we treat every fact as a function mapping variables to values, and we remove a fact $f_1$ from the relation of $\varphi$ if there exists another atom containing $w$ that does not contain a fact that agrees with $f_1$ on their common variables. Notice that this filtering does not affect the query answers, and that the other atoms that contain $w$ are now no longer needed: the query will have the same answers if they are removed.

Consider the CQ  $\Qcount(\vec{x}, w)$ with the same body as $\Qfull$ but without all atoms that contain $w$, and with the new atom $\varphi'$, where the relation for $\varphi'$ is obtained from the filtered $\varphi$ by replacing the $w$ values with their counts. That is, we sort $\varphi$ according to all variables other than $w$, then scan the relation and replace every set of tuples that agree on the other variables with one tuple where the $w$ value is the number of tuples we removed in this replacement.

Now we can use the known algorithm from \Cref{thm:known-dichotomy} for free-connex CQs with no disruptive trios to solve $\Qcount$. As we argued that all the transformations above maintain the query answers, this algorithm results in exactly $Q(D)$. We also notice that these transformations can be done in $O(|D|\log|D|)$ time.

Next, we prove the negative side of the dichotomy. That is, we assume that $Q'$ is either not free-connex or contains a disruptive trio and prove the hardness of $Q$. 
For this, we use the following characterization of free-connex CQs.
Given a CQ $Q$, a {\em free-path} is a chordless path $(x_1,y_1,\ldots,y_k,x_2)$ in $H(Q)$ with $k\geq 1$, such that $x_1,x_2\in \free(Q)$, and $y_1,\ldots,y_k\not\in \free(Q)$.
Here, a {\em chordless path} is a simple path (i.e., a sequence of distinct vertices where every two consecutive vertices co-occur in at least one hyperedge) where no two non-consecutive vertices are neighbors.
An acyclic CQ $Q$ is free-connex if and only if it does not contain a free-path~\cite{10.1007/978-3-540-74915-8_18}.

Consider the CQ $Q''(\vec{x})$ with the same body as $Q$ and $Q'$, but where we removed the aggregated value from the query head.
The first case is that $Q''$ is not free-connex or has disruptive trios. In this case, according to \Cref{thm:known-dichotomy}, direct access for $Q''$ is not in $\linlog$. Since each answer to $Q$ contains an answer to $Q''$ and the answers have the same order, we can use $Q$ to solve $Q''$. So, we conclude that $Q$ has no direct access in $\linlog$ either.
The second case is that $Q''$ is free-connex with no disruptive trios.
If $Q'$ has a disruptive trio, its last variable has to be $w$ since $Q''$ has no disruptive trios. In other words, the disruptive trio comprises of $\{x_i,x_j,w\}$ where $x_i,x_j\in\vec{x}$ and $(x_i,w,x_j)$ is a chordless path. This chordless path constitutes a free-path in $Q''$, contradicting the fact that $Q''$ is free-connex. So, we conclude that $Q'$ has no disruptive trios.
This means that $Q'$ is acyclic (since $Q''$ is acyclic and they have the same body) but not free-connex. Thus, $Q'$ has a free-path that is not a free-path in $Q''$, and so it must use $w$.
We conclude that this path is of the form $(x_i,v_1,\ldots,v_k,w)$ with $x_i\in\vec{x}$ and $v_1,\ldots,v_k\in\vec{y}$.
Since all queries we discussed have the same body, this chordless path also appears in $Q$. 

Since $(x_i,v_1,\ldots,v_k,w)$ is a chordless path, the \ACQ $Q_{|\{x_i,v_1,\ldots,v_k,w\}}$ is structurally equivalent to 
$$Q_H^k(x, \aggcountd(w)) \datarule R_0(x,y_1), R_1(y_1,y_2),\ldots,R_{k-1}(y_{k-1},y_k), R_k(y_k,w)\,$$ 
where $x_i$, $v_j$, and $w$ take the roles of $x$, $y_j$, and $w$ respectively, for all $1\leq j\leq k$. \Cref{lem:hsc} determines that $Q_H^k$ has no direct access in $\linlog$, assuming \HSC.
Using \Cref{lem:reduction}, we conclude that direct access for $Q$ is not in $\linlog$.
\end{proof}

The following example illustrates the construction in the positive side of the proof.
\begin{exa}\label{example:theorem-countd}
Consider the following \ACQ:
 $$Q(x_1,x_2,x_3,\aggcountd(w))\datarule R(x_1,w,y),R(x_1,x_2,w),S(x_2,x_3)$$
 For this query we construct $$\Qfull(x_1,x_2,x_3,w)\datarule R_1(x_1,w),R_2(x_1,x_2,w),S(x_2,x_3)\,,$$ which is self-join-free query and without existential variables, by replacing $R$ with unique names and removing $y$. We then get $$\Qcount(x_1,x_2,x_3,w)\datarule R_2'(x_1,x_2,w),S(x_2,x_3)$$ by keeping only the maximal atom containing $w$.
Given $R$ and $S$ as follows, the following relations will be built. $R_1$ is obtained from $R$ by projecting away the third column. $R_2'$ is obtained by filtering $R_2$ according to $R_1$ (which removes the fact $R_2(a_1,b_2,b_3)$), then grouping by the first two columns and counting.
\begin{center}
\begin{tabular}[t]{cc}
\multicolumn{2}{l}{$S$}\\\toprule
$b_1$ & $c_1$ \\
$b_2$ & $c_1$ \\
$b_2$ & $c_2$ \\
\bottomrule
\end{tabular}\quad\quad
\begin{tabular}[t]{ccc}
\multicolumn{3}{l}{$R=R_2$}\\\toprule
$a_1$ & $b_1$ & $b_1$ \\
$a_1$ & $b_1$ & $b_2$ \\
$a_1$ & $b_2$ & $b_1$ \\
$a_1$ & $b_2$ & $b_3$ \\
$a_2$ & $b_2$ & $b_2$ \\
\bottomrule
\end{tabular}\quad\quad
\begin{tabular}[t]{cc}
\multicolumn{2}{l}{$R_1$}\\\toprule
$a_1$ & $b_1$ \\
$a_1$ & $b_2$ \\
$a_2$ & $b_2$ \\
\bottomrule
\end{tabular}\quad\quad
\begin{tabular}[t]{ccc}
\multicolumn{3}{l}{$R_2'$}\\\toprule
$a_1$ & $b_1$ & $2$ \\
$a_1$ & $b_2$ & $1$ \\
$a_2$ & $b_2$ & $1$ \\
\bottomrule
\end{tabular}
\end{center}
\end{exa}

\begin{rem}\label{rem:countd-small-domain}
\Cref{thm:countd-dichotomy} shows which \ACQs belong to the positive side of \Cref{cor:aggregate-tractable}, but become intractable when the aggregation is replaced with count-distinct.
    Nevertheless, it is important to observe that there are cases where \Cref{thm:annotated-dichotomy} can be used to compute \ACQs with count-distinct even for the intractable cases of \Cref{thm:countd-dichotomy}: when the size of the domain $\cntdomain$ of the distinct elements we count is bounded by a logarithm in the input size $|D|$. 
    Such an assumption can be realistic when we count, say, distinct categories from a small ontology (e.g., item categories in a sales context), distinct countries from a small collection of countries, and so on. 
    In such cases, we can compute the exact set of distinct elements, and not just their count, by using the set semiring $(\powerset(\cntdomain),\cup, \cap, \varnothing, \cntdomain)$ since the operations $\cup$ and $\cap$ can be performed in logarithmic time for logarithmic domains.\qed
\end{rem}

\section{Incorporating the Annotation and Aggregation in the Order}\label{sec:inside}

The results of the previous section apply when the lexicographic order does not include the computed value,  or equivalently when the computed value is last in the head of the query. In this section, we explore the ability to include the computed value earlier in the lexicographic order. To this end, we assume that the underlying commutative semiring has an ordered domain. When the domain is numerical, we will implicitly assume the natural order (i.e., smaller numbers precede larger numbers) without mentioning it. In terms of the computational model, we assume that we can compare two given elements of the domain in time logarithmic in the size of the input. 

Let $(\K,\oplus,\otimes,\zero,\one)$ be a commutative semiring with an underlying order $\succeq$ over $\K$. The semiring is said to be \e{$\otimes$-monotone} if for every $c \in \K$ the function $f_c:\K \rightarrow \K$, defined by $f_c(y) = c \otimes y$, is monotone. This means that either $c\otimes a\succeq c\otimes b$ whenever $a\succeq b$, or  $c\otimes a\succeq c\otimes b$ whenever $b\succeq a$. 
Computationally, we assume that we can determine efficiently (in logarithmic time in the input) whether a given $c$ is such that the function $f_c(x)=c\otimes x$ is non-decreasing or non-increasing. All specific semirings that we mention in the paper are $\otimes$-monotone. 
Our main result for this section is the following dichotomy.
Recall that the \THREESUM  conjecture~\cite{3SUM-GAJENTAAN1995165,3SUM-Patrascu} states
it takes $N^{2-o(1)}$ time to determine whether a given set of $N$ elements from $\set{-N^3,\ldots,N^3}$ contains distinct elements $a,b,c$ such that $a+b=c$.

\begin{thm}\label{thm:dichotomy-general-annotation}
Let $(\K,\oplus,\otimes,\zero,\one)$ be a $\otimes$-monotone logarithmic-time commutative semiring, and
$Q(\vec x,\star,\vec z)$ a \cqstar.
\begin{enumerate}
    \item If $Q$ is free-connex with no disruptive trio and there exists an atom of $Q$ containing $\neigh(\vec z) \cap \free(Q)$,  then direct access for $Q$ is in $\linlog$. 
    \item Otherwise, if $Q$ is also self-join-free and $(\K,\oplus,\otimes,\zero,\one)$ is one of the counting, numerical, max tropical, or min tropical semirings, then direct access for $Q$ is not in $\linlog$, assuming the
    \THREESUM, \sparseBMM, and \HYPERCLIQUE hypotheses.
\end{enumerate}
\end{thm}

Recall that the notation $\neigh(\vec z)$ contains also all variables of $\vec z$.
Before we prove \Cref{thm:dichotomy-general-annotation}, we illustrate the tractability condition with a simple special case: If $Q(\vec x,\star,\vec z)$ is free-connex and all variables of $\vec z$ act ``as a single variable,'' then direct access for $Q$ is in $\linlog$. More formally:

\begin{cor}\label{cor:general-annotations-order-tractability}
Let $(\K,\oplus,\otimes,\zero,\one)$ be a $\otimes$-monotone logarithmic-time commutative semiring, and
$Q(\vec x,\star,\vec z)$ a free-connex \cqstar with no disruptive trio. If every atom of $Q$ contains either all variables of $\vec z$ or none of them, then direct access for $Q$ is in $\linlog$.
\end{cor}

\begin{proof}
    We first use \Cref{cor:pos-existential-removal} to eliminate the existential variables of $Q$ and work with a full acyclic \cqstar $\Qpfree(\vec{x},\star,\vec{z})$.
    We notice that $\Qpfree$, like $Q$, is such that every atom contains either all variables of $\vec{z}$ or none of them and there is no disruptive trio.
    If $\vec z$ is empty, then every atom contains $\neigh(\vec z)$.
    Otherwise, denote the last variable in $\vec z$ as $z_\last$. It has been established that a full acyclic CQ with no disruptive trios has an atom that contains the last free variable and all of its neighbors~\cite[proof of Lemma 3.9]{carmeli2023tractable};
    hence, $\Qpfree$ has an atom $\varphi$ that contains $z_\last$ and all of its neighbors. 
    Since each atom of $Q$ contains either all or none of the variables of $\vec z$, we know that $\neigh(\vec z)=\neigh(z_\last)$,
    hence $\varphi$ contains every variable in $\neigh(\vec z)$. From \Cref{thm:dichotomy-general-annotation} we can now conclude that direct access for $\Qpfree$ is in $\linlog$.
\end{proof}

For example, \Cref{cor:general-annotations-order-tractability} (and \Cref{thm:dichotomy-general-annotation}) imply that
direct access for the following \cqstar is in $\linlog$ over databases annotated with the numerical semiring.
\begin{equation*}
   Q(x_1,x_2,\star,z_1,z_2) \datarule  R(x_1,x_2), S(x_2,z_1,z_2), T(z_1,z_2)\label{eq:almost-last-example}
\end{equation*}
The variables that follow the computed value $\star$ are $z_1$ and $z_2$, and indeed, every atom either contains both $z_1$ and $z_2$ (as the second and third atoms) or contains none of them (as the first atom). Hence, direct access is tractable according to \Cref{cor:general-annotations-order-tractability}. Similarly, the atom $S$ contains $\neigh(\vec z)$
(i.e., the variables $x_2, z_1, z_2$), so \Cref{thm:dichotomy-general-annotation} also identifies this query as tractable.

However, \Cref{cor:general-annotations-order-tractability} fails to capture the tractability of the following example.
\begin{equation*}
   Q(x_1,x_2,\star,z_1,z_2) \datarule  R(x_1,x_2), S(x_2,z_1,z_2), T(z_2)\label{eq:almost-last-example-not-by-cor}
\end{equation*}
The variables $z_1$ and $z_2$ do not occur jointly in all atoms, as only $z_2$ appears in $T(z_2)$. Yet, there is an atom that contains both $z_1$ and $z_2$, so direct access for $Q$ is in $\linlog$ according to \Cref{thm:dichotomy-general-annotation}.

As a negative example, consider the simplest possible Cartesian product \cqstar 
$R\times S$ for unary $R$ and $S$, where we wish to have the annotation \e{first} in the order.
\begin{equation}\label{query:intractable-annotated-order} 
\Qstartimes(\star,x, y) \datarule R(x),S(y)\,.
\end{equation}
This \cqstar falls in the hard side of \Cref{thm:dichotomy-general-annotation}, since there is no atom that contains both $x$ and $y$. In contrast, \Cref{thm:dichotomy-general-annotation} implies that whenever $\star$ is not first, as is the case with $(x,\star,y)$, the query is tractable. 

In the remainder of this section, we prove \Cref{thm:dichotomy-general-annotation}. We begin with the tractability side (Section~\ref{sec:general:tractability}), then complete the proof with the hardness side (Section~\ref{sec:general:hardness}), and discuss the implications for \ACQs (Section~\ref{sec:general:acqs}).

\subsection{Tractability Side of Theorem~\ref{thm:dichotomy-general-annotation}}\label{sec:general:tractability}
The following lemma states the tractability side of \Cref{thm:dichotomy-general-annotation}. The proof includes the algorithm.

\begin{lem}\label{lemma:general-annotations-order-tractability}
Let $(\K,\oplus,\otimes,\zero,\one)$ be a $\otimes$-monotone logarithmic-time commutative semiring, and
$Q(\vec x,\star,\vec z)$ a free-connex \cqstar with no disruptive trio. If $Q$ has an atom that contains $\neigh(\vec z) \cap \free(Q)$, then direct access for $Q$ is in $\linlog$.
\end{lem}

\begin{proof}
    The proof idea is as follows. After eliminating the projection (\Cref{lemma:existential-elimination-semiring-Qfree}), an answer is obtained by joining a selected fact from each relation, and its annotation is obtained by the multiplication of the annotations of the selected facts.
    Denote by $A\before$ the set of atoms that contain only variables of $\vec x$, and by $A\after$ the set of remaining atoms.
    The annotation of each answer can be expressed as a product $c\before \otimes c\after$, where $c\before$ is determined once $\vec x$ is assigned values, and $c\after$ is determined once the remaining variables are assigned values. Once $c\before$ is determined, sorting by $\star = c\before \otimes c\after$ is the same as sorting by $c\after$. 
    Let $\varphi$ be an atom that contains $\neigh(\vec z) \cap \free(Q)$. 

    We show that we can extend $\varphi$ with a fresh variable, assign this variable with the annotation $c\after$, and then use a known direct-access solution on the extended database and query. We explain this idea in the remainder of the proof.

    We first eliminate the existential variables of $Q$ to work with a full acyclic \cqstar  $\Qpfree(\vec{x},\star,\vec{z})$. According to \Cref{lemma:existential-elimination-semiring-Qfree}, a $\K$-database $(D, \tau)$ can be transformed in $O(|D| \log |D|)$ time to a $\K$-database $(D_\full, \tau_\full)$ such that $Q(D, \tau) = \Qpfree(D_\full, \tau_\full)$.
    We notice that the obtained $\Qpfree$, like $Q$, has an atom $\varphi_\full$ that contains $\neigh_Q(\vec z) \cap \free(Q) = \neigh_{\Qpfree}(\vec z)$. 
    Every atom in $A\after$ contains at least one variable in $\vec z$,
    thus  $\varphi_\full$ contains every variable that occurs in any atom in $A\after$.
    
    Next, we add a new variable $y$ to $\Qpfree$:
    \begin{itemize} 
        \item We add $y$ to the atom $\varphi_\full$;
        \item Replace the head $(\vec{x},\star,\vec{z})$ with 
        $(\vec{x},y,\vec{z},\star)$.
    \end{itemize}
    We denote the resulting \cqstar by $\Qpfree^{+y}$.

    We first prove that direct access for $\Qpfree^{+y}$ is in $\linlog$. Due to \Cref{thm:annotated-dichotomy}, it suffices to show that $\Qpfree^{+y}$ is acyclic and that it does not contain a disruptive trio. To see that $\Qpfree^{+y}$ is acyclic, we can use the same join tree as $\Qpfree$;  the running-intersection property is preserved since we add $y$ to only one atom. Assume, by way of contradiction, that there is a disruptive trio. As $\Qpfree$ has no disruptive trio, and the variable sequence of $\Qpfree^{+y}$ differs from that of  $\Qpfree$ by $y$ alone, any disruptive trio in $\Qpfree^{+y}$ must include $y$. 
    
    Suppose that $y$ forms a disruptive trio together with $x_a$ and $x_b$. Recall that $y$ was added to only one atom, namely $\varphi_\full$, and so all of its neighbors are neighbors of each other. Therefore, the only possibility for $\set{y,x_a,x_b}$ to form a disruptive trio is the following:   
    \begin{itemize}       
        \item $x_a$ and $x_b$ are neighbors in an atom $\varphi'$;
        \item $x_a$ and $y$ are non-neighbors;
        \item $x_b$ and $y$ are neighbors in $\varphi_\full$;
        \item $x_b$ appears after $y$ and $x_a$ in the head of $\Qpfree^{+y}$. 
    \end{itemize}
    Since $x_b$ appears after $y$, we know that $x_b$ is in $\vec{z}$. Hence, $\varphi'$ is in $A\after$. As $\varphi_\full$ contains the variables of all atoms in $A\after$, it also contains $x_a$, contradicting the fact that $x_a$ and $y$ are non-neighbors. 
    In conclusion, direct access for $\Qpfree^{+y}$ is in $\linlog$, as claimed.

    Next, we show how we can use direct access to $\Qpfree^{+y}$ to obtain direct access for $\Qpfree$.
    Given a $\K$-database $(D_\full, \tau_\full)$ for $\Qpfree$, we start by modifying the $\K$-database to obtain a new $\K$-database $(D_\full^{+y}, \tau_\full^{+y})$ for $\Qpfree^{+y}$.
    Let $R_\varphi$ be the relation symbol of $\varphi_\full$. 
    For any $R_\varphi$-fact $f$ in $D_\full$, and for every atom $\varphi'$ in $A\after$, if ${D_\full}$ does not contain a fact $f'$ such that $f$ and $f'$ agree w.r.t.~$\varphi_\full$ and $\varphi'$, then we can ignore $f$.
    Also, recall that $\varphi_\full$ contains all variables that occur in $A\after$. Hence, for every $R_\varphi$-fact $f$ of $D_\full$ and atom $\varphi'\in A\after$, there is at most one fact $f'$ of $\varphi'$ that agrees with $f$. 
    We set the value of $y$ for every $R_\varphi$-fact $f$ of $D_\full^{+y}$ to be the product of the annotations of the facts that agree with $f$ in the relations of atoms in $A\after$. 
    Since $|A\after|$ is a constant in terms of data complexity, we can compute the product in constant time per fact, and the construction takes linear time.
    The annotations of facts of $D_\full^{+y}$ will be identical to the annotations of the corresponding facts of $D_\full$ as the annotations are unchanged.
    
    We apply the direct-access algorithm on $D_\full^{+y}$ twice: once normally, and once when reversing the order over the domain of $y$. This gives us two answers, $\vec{a}$ and $\vec{a}_r$, that have the same assignment to $\vec{x}$. We compute the result of $c\before$ of these answers and check whether multiplying by ${c\before}$ is monotone non-decreasing or monotone non-increasing (which we assume is doable in logarithmic time). 
    If it is monotone non-decreasing, then sorting by $c\after$ is identical to sorting by $c\before \otimes c\after$ and we choose $\vec{a}$ as the answer. If it is monotone non-increasing, then sorting by $c\after$ in reverse order is identical to sorting by $c\before \otimes c\after$, and we choose $\vec{a}_r$ as the answer. In either case, we return the answer after projecting out variable the value corresponding to $y$.\footnote{It is possible to only use one direct-access structure and support accessing a variable in both orders using a suffix-sum in addition to a prefix-sum at preprocessing. During access, we first determine the value for the first $i$ variables, and then decide in which direction to access the next variable. This would give a practical improvement but the same complexity. Since we do not give the details of the direct access algorithm for CQs in this paper, for the sake of simplicity of arguments, we use the algorithm twice in the proof instead.}
    Hence, we get direct access for $\Qpfree$ in $\linlog$, and so, direct access for $Q$ in $\linlog$, as required.
\end{proof}

\subsection{Hardness Side of Theorem~\ref{thm:dichotomy-general-annotation}}\label{sec:general:hardness}
To prove the hardness side of the theorem, we
first show hardness of the \cqstar $\Qstartimes$ of~\eqref{query:intractable-annotated-order}.
The next lemma states that under the \THREESUM conjecture, direct access for $\Qstartimes$ is impossible over $\K$-databases with semirings that gave positive results in the previous section.

\begin{lem}\label{lem:X}
Let $(\K,\oplus,\otimes,\zero,\one)$ be one of the counting, numerical, max tropical, or min tropical semirings.
Direct access for the \cqstar $\Qstartimes$ (of~\eqref{query:intractable-annotated-order}) 
is not in $\linlog$ over $\K$-databases, assuming the \THREESUM conjecture.
Moreover, this holds even if the annotation is in the domain $[1,|D|^4 - 1]$.
\end{lem}

\begin{proof}

We will show a reduction from the \THREESUM problem. Given a set $V$ of $N$ values over the universe $\set{-N^3, \ldots, N^3}$, we construct a logarithmic-time, homomorphism $h:\Z\rightarrow\K$, which we define later on, such that for every $a,b\in V$ we get $h(a+b) = h(a) \otimes h(b)$.
The $\K$-relation $(D,\tau)$ that we construct consists of the facts $R(a)$ and $S(a)$ for every $a\neq 0$. As for the annotation, we define $\tau(R(a))=\tau(S(a))=h(a)$. Hence, the annotation of an answer $(a,b)$ is $h(a) \otimes h(b)$.

Now, assume that we have efficient direct access to $\Qstartimes(D)$.
For every value $c \in V$, we compute $h(c)$ and use binary search on $Q(D)$ to find an answer $(a, b)$ with the annotation $h(c)$. If we find such an answer, this means that there exists some $a,b \in V$ such that $h(a) \otimes h(b) = h(c)$. We will show that $h$ is injective within a subset of $\Z$ that contains $a$, $b$, and $a+b$ for all $a,b \in V$. As a result, we get that $h(a) \otimes h(b) = h(c)$ implies $a + b = c$.
If $a\neq b$, then we have found the required triple in $V$, and we are done. If $a = b$, then we check whether the preceding or the following answer also has the annotation $h(c)$. If so, then we have found our triple since at most one of these answers satisfies $a=b$ (i.e., $a=b=c/2$). If we do not find the triple we seek after iterating over $V$, then we conclude that the answer to the \THREESUM problem is negative.

As there are $N^2$ answers to $\Qstartimes$, binary search for an answer by the annotation requires $O(\log N)$ access calls. Given a value $c \in V$, using logarithmic-time direct access we can find an answer, if exists, in $O(\log^2 N)$ time. In total, we handle all values in $V$ and solve the \THREESUM problem in $O(N \cdot \log^2 N)$ time, contradicting the \THREESUM conjecture. 

The rest of this proof is dedicated to finding a suitable logarithmic-time homomorphism $h$, given $V$. When discussing the min tropical or max tropical semiring, we can simply use the identity as the homomorphism since $\otimes$ is the numerical addition ($+$). As for the numerical and counting semiring, $\otimes$ is defined as $\cdot$, so our homomorphism is more involved. We handle these in the remainder of the proof.

In the following paragraphs, we make use of standard notations of number theory (see, e.g.,~\cite{DBLP:books/lib/Cohen93}). We denote by $(\Z/m\Z)$, or $\Z_m$ for short, the group of integers modulo $m$ with the addition ($+$) operator. We denote with $(\Z/m\Z)^*$, or $\Z^*_m$ for short, the group of integers modulo $m$ with the multiplication ($\cdot$) operator. Recall that $|\Z^*_m| = \phi(m)$ where $\phi$ stands for the Euler's totient function.

To define the homomorphism $h$, we choose a value $m_0$ with the following properties:
\begin{enumerate}
    \item $\phi(m_0) \geq 4N^3+1$.\label{item:m-greater-than}
    \item The multiplicative modular group $\Z^*_{m_0}$ has $2$ for a generator.\label{item:2-generator}
    \item $\log m_0 = O(\log N)$.\label{item:m-small-enough}
\end{enumerate}
We later show how we use these properties of $m_0$ to define $h$. We choose $m_0 = 3^{\lceil \log_3 (4N^3+1)\rceil + 1}$, and before we proceed, we will prove that indeed all three properties hold for this $m_0$. 
By the definition of the totient function, $\phi(m_0) = 2 \cdot 3^{\lceil \log_3 (4N^3+1)\rceil} \geq 4N^3+1$, so \Cref{item:m-greater-than} holds. 
Since $m_0$ is a power of the prime $3$, it holds that $Z_{m_0}^*$ is cyclic \cite[Theorem 1.4.1]{DBLP:books/lib/Cohen93}. In addition, the number $2$ is a generator for each power of $3$ because
$2$ is a generator for $Z_{3^2}^*$~\cite[Lemma 1.4.5]{DBLP:books/lib/Cohen93}. Hence, \Cref{item:2-generator} holds. 
Lastly, $m_0 = 3^{\lceil \log_3 (4N^3+1)\rceil + 1} \leq 9 \cdot 3^{\log_3 (4N^3+1)} \leq 9 \cdot (4N^3+1)$, and so $\log m_0 = O(\log N)$ and \Cref{item:m-small-enough} holds.

It remains to construct $h: \Z \rightarrow \K$ using $m_0$, and we do so as follows.
Define $h\defeq h_3\circ h_2\circ h_1$ 
(that is, $h(a)=h_3(h_2(h_1(a)))$) for the following homomorphisms $h_1$, $h_2$ and $h_3$:
\begin{itemize}
    \item $h_1: \Z \rightarrow \Z_{\phi(m_0)}$ is defined by $h_1(a) = a \pmod{\phi(m_0)}$. For every $a,b \in V$, we know that $a$, $b$, and $a+b$ are all in the range of $[-2N^3, \ldots, 2N^3]$. Due to \Cref{item:m-greater-than}, we have $\phi(m_0) \ge 4N^3+1$, and therefore $h_1$ is injective for any possible $a,b,a+b$ originating from $a,b\in V$.
    \item $h_2:\Z_{\phi(m_0)} \rightarrow \Z^*_{m_0}$ is defined as follows. Due to \Cref{item:2-generator}, the number $2$ is a generator for $\Z^*_{m_0}$, and so we can use as $h_2$ the well-known isomorphism between modular additive groups and cyclic groups: $h_2(a) = 2^a \pmod{m_0}$.
    \item $h_3: \Z^*_{m_0} \rightarrow \K$ is simply the identity function. Note when discussing the numerical semiring, $\K = \Q$, and when discussing the counting semiring, $\K = \N$. Therefore, $h_3$ is well defined in the case of each of our semirings.
\end{itemize}
 Note that $h_1$ is injective over the range of possible values for $a, b, a+b$ for $a,b \in V$, and $h_2$ and $h_3$ are injective in their entire domain. As a composition, for every possible value $a,b \in V$ the function $h$ is injective and is a homomorphism: $h(a+b)=h(a)\otimes h(b)$. In addition, we can compute $h(a)$ in logarithmic time: modular exponentiation in $\Z^*_{m_0}$ can be done using $O(\log m_0)$ multiplications, so the computation of $h$ can be done in $O(\log m_0)$ time. Due to \Cref{item:m-small-enough}, $\log m_0 = O(\log N)$, and therefore, $h_1$, $h_2$, and $h_3$ all take logarithmic time. In addition, due to \Cref{item:m-small-enough}, the values computed by $h$ can be represented in a logarithmic number of bits and fit in the registers of our RAM model.

 Lastly, recall that the values $h$ outputs are the values used as annotations in the database. Since the output of $h_2$ is in $\Z^*_{m_0}$, and $m_0=3^{\lceil \log_3 (4N^3+1)\rceil + 1}$, then the output of $h_2$ is in the range of $[1, \ldots, 12N^3+2]$. As $h_3$ is defined as the identity, we get that the output of $h$ is in that same range, $[1, \ldots, 12N^3+2]$. For every $N \geq 13$, the output of $h$ is in the range of $[1, \ldots, N^4 - 1]$.
\end{proof}

Next, we build on the hardness in \Cref{lem:X} to reason about a more general case.

\begin{lem}\label{lemma:general-annotations-dichotomy-case-1}
Let $(\K,\oplus,\otimes,\zero,\one)$ be one of the counting, numerical, max tropical, or min tropical semirings and let $Q(\vec x,\star,\vec z)$ be a full, acyclic, self-join-free \cqstar with no disruptive trio.
If direct access for $Q$ over $\K$-databases is in $\linlog$, then all variables in $\vec z$ neighbor each other, assuming the \THREESUM conjecture.
\end{lem}

\begin{proof}
We will prove the contrapositive of the lemma: we assume that there are two variables $z_1, z_2$ in $\vec z$ that are not neighbors and show that direct access for $Q$ is not in $\linlog$.
More specifically, we show how to use direct access for $Q$ to obtain direct access for $\Qstartimes(\star,x, y) \datarule R(x),S(y)$. 
Given a $\K$-database $(D_\star, \tau_\star)$ for $\Qstartimes$, we show how to construct a $\K$-database $(D, \tau)$ for $Q$ such that $Q(D,\tau) = \Qstartimes(D_\star, \tau_\star)$. 
Since by \Cref{lem:X} direct access for $\Qstartimes$ is not in $\linlog$, we get that neither is direct access for $Q$.

Since $z_1$ and $z_2$ are not neighbors, the atoms of $Q$ can be partitioned into those containing $z_1$, those containing $z_2$, and those containing none of them. Denote the sets of these atoms by $\varphi_{z_1}$, $\varphi_{z_2}$, and $\varphi_\varnothing$, respectively.
For every fact $R(c)$ in $D_\star$ we add for every atom $\varphi$ in $\varphi_{z_1}$ a fact $f_{c,\varphi}$ to $D$ derived from $\varphi$ by replacing $z_1$ with $c$ and every other variable with an arbitrary value $a_0$. We then choose an arbitrary atom $\varphi'$ in $\varphi_{z_1}$ and define $\tau(f_{c,\varphi'}) = \tau_\star(R(c))$, and for every other atom $\varphi$ in $\varphi_{z_1}$ we define $\tau(f_{c,\varphi}) = \one$.
Similarly, for every fact $S(c)$ in $D_\star$ we add for every atom $\varphi$ in $\varphi_{z_2}$ a fact $f_{c,\varphi}$ to $D$ derived from $\varphi$ by replacing $z_1$ with $c$ and every other variable with $a_0$. We then choose an arbitrary atom $\varphi'$ in $\varphi_{z_2}$ and define $\tau(f_{c,\varphi'}) = \tau_\star(S(c))$, and for every other atom $\varphi$ in $\varphi_{z_2}$ we define $\tau(f_{c,\varphi}) = \one$.
For every atom $\varphi$ in $\varphi_\varnothing$ we add a single fact derived from $\varphi$ by replacing every variable with $a_0$, annotated by $\one$.

Now note that there is a bijection from $Q(D,\tau)$ to $\Qstartimes(D_\star, \tau_\star)$ mapping $(c_1, c_2)$ to a tuple of the form $(a_0,..., c_1,...,c_2,...a_0)$ where $c_1$ and $c_2$ are in place of $z_1$ and $z_2$ in the head of $\Qstartimes$. Moreover, $\tau_\star((c_1, c_2)) = \tau((a_0,..., c_1,...,c_2,...a_0))$. This means that we can use $Q$, $D$ and $\tau$ to provide direct access for $\Qstartimes$, $D_\star$ and $\tau_\star$. As $|Q|$ is constant and we constructed $D$ in $|D_\star||Q|$ time, we get direct access for $\Qstartimes$ in $\linlog$ time.
\end{proof}

We also build on the hardness of queries with disruptive trios, as stated in \Cref{thm:known-dichotomy}, to deduce the hardness in case there is a variable before the annotation that neighbors some but not all variables after the annotation.

\begin{lem}\label{lemma:general-annotations-dichotomy-case-2}
Let $Q(\vec x,\star,\vec z)$ be a full, acyclic, self-join-free \cqstar with no disruptive trio such that all variables in $\vec z$ neighbor each other.
If direct access for $Q$ over $\K$ databases is in $\linlog$, then every variable $x \in \vec x$ that neighbors a single variable in $\vec z$ neighbors all variables in $\vec z$, assuming the \sparseBMM hypothesis.
\end{lem}

\begin{proof}
    We will prove the contrapositive of the lemma. We assume that there are variables $x \in \vec x$ and $z_i,z_j \in \vec z$ 
    such that $x$ neighbors $z_j$ but not $z_i$, and show that direct access for $Q$ is not in $\linlog$.
    Let $Q'$ be the CQ $Q'(\vec x,z_i,\vec z)$ with the same body as $Q$. Note that $z_i$ appears twice in the head of $Q'$, and therefore the second appearance does not affect the order.
    From \Cref{thm:known-dichotomy}, since $Q$ is acyclic and $x$, $z_i$ and $z_j$ form a disruptive trio, we know that direct access for $Q'(\vec x,z_i,\vec z)$ is not in $\linlog$, assuming the \sparseBMM hypothesis.
      Therefore, if we show that we can use direct access for $Q$ to obtain direct access for $Q'$, then we get that direct access for $Q$ is not in $\linlog$.

    Given a database $D'$ for $Q'$, we show how to construct a $\K$-database $(D, \tau)$ for $Q$.
    We begin by performing the Yannakakis full reduction to eliminate dangling tuples~\cite{DBLP:conf/vldb/Yannakakis81}. 
    We can then assume that every fact $f$ of $D$ is consistent with $Q$ in the sense that we can obtain $f$ from the corresponding atom of $Q$ by assigning some values to the variables; we then denote by $f[x]$ the value of $f$ in the positions of the variable $x$.
    We then construct $D$ by first adding every fact in $D'$ to $D$. Then we choose a single arbitrary relation in $D$ that corresponds to an atom that contains $z_i$, and for every fact $f$ in the relation, we define $\tau(f) = f[z_i]$. For every other fact $f$ in $D$ we define $\tau(f) = \one$.
    
    Since the bodies of $Q'$ and $Q$ are identical, there is a bijection from $Q'(D')$ to $Q(D, \tau)$ mapping $f'$ to $f$ such that $f[w] = f'[w]$ for every variable, where $\tau(f)=f[z_i]$.
    Therefore, ordering $Q(D, \tau)$ by $\vec x,\star,\vec z$ is the same as ordering $Q'(D')$ by $\vec x,z_i,\vec z$.
    For any database $D'$, we can obtain direct access for $Q'(D')$ by producing the equivalent database $(D, \tau)$ and then providing direct access for $Q(D, \tau)$.
\end{proof}

We will use the following lemma, often seen as folklore \cite[Lemma 5.4]{carmeli2023tractable} \cite[Proposition 5.3]{carmeli2023efficientlyenumeratingminimaltriangulations}. For completeness, we offer a proof suitable for our formalization.
\begin{lem}\label{lemma:atom-neighbors}
Let $Q$ be an acyclic CQ and let $X$ be a set of variables of $Q$.
If all pairs of variables of $X$ are neighbors in $Q$, then there exists an atom $\varphi$ in $Q$ such that $X \subseteq \var(\varphi)$.
\end{lem}

\begin{proof}
Let $T$ be a join tree of $Q$. Due to the running-intersection property, for each variable $x_i$ in $X$, the induced subgraph of vertices containing $x_i$ is a (connected) subtree, and let us denote it by $T_{x_i}$. It is known that every set of subtrees of $T$ fulfills the \e{Helly property}~\cite{DBLP:journals/networks/Trotter83}: if the intersection of every two subtrees in the set is non-empty, then the intersection of all subtrees in the set is non-empty. Since every two variables $x_i$ and $x_j$ in $X$ are neighbors, the subtrees $T_{x_i}$ and $T_{x_j}$ share some node of $T$. Therefore, the intersection of all subtrees contains some node of the join tree. That means that there is an atom which induced that node, whose variables contain $X$.
\end{proof}

We can now prove the negative side of the dichotomy for the case of free-connex \cqstars{} with no disruptive trios.

\begin{lem}\label{lemma:general-annotations-order-dichotomy-full-negative}
Let $(\K,\oplus,\otimes,\zero,\one)$ be the counting, numerical, max tropical, or min tropical semirings, and
$Q(\vec x,\star,\vec z)$ a free-connex, self-join-free \cqstar with no disruptive trio. If direct access for $Q$ is in $\linlog$, then there exists an atom of $Q$ that contains $\neigh(\vec z) \cap \free(Q)$, assuming the \THREESUM and \sparseBMM hypotheses.
\end{lem}

\begin{proof}
    Since direct access for $Q$ is in $\linlog$, by \Cref{thm:existential-removal} we know that direct access for $\Qpfree$ is in $\linlog$. Since $\Qpfree$ contains all free variables of $Q$ and every atom of $\Qpfree$ is contained in an atom of $Q$, it is enough to show that there exists an atom in $\Qpfree$ that contains $\neigh(\vec z)$. Let us consider $\Qpfree$ from now on.
    As direct access for $\Qpfree$ is in $\linlog$, by \Cref{lemma:general-annotations-dichotomy-case-1} we know that all variables in $\vec z$ neighbor each other. By \Cref{lemma:general-annotations-dichotomy-case-2}, we also know that every variable in $\vec x$ that neighbors a single variable in $\vec z$ neighbors all variables in $\vec z$.

    Let $x_1, x_2 \in \vec x$ be two neighbors of $\vec z$. We claim that $x_1$ neighbors $x_2$.
    Let $z_1$ be an arbitrary variable in $\vec z$. As $x_1$ and $x_2$ are in $\neigh(\vec z)$ we know that both $x_1$ and $x_2$ neighbor all variables in $\vec z$ and in particular $z_1$. Both $x_1$ and $x_2$ neighbor $z_1$, and $z_1$ appears after $x_1$ and $x_2$ in the order of $\Qpfree$. Since we know there are no disruptive trios, it must mean that $x_1$ neighbors $x_2$.

    In conclusion, let $y_1, y_2$ be two variables in $\neigh(\vec z)$. 
    If $y_1, y_2$ are in $\vec z$, they neighbor each other as we concluded from \Cref{lemma:general-annotations-dichotomy-case-1}.
    If $y_1 \in \vec z$ and $y_2 \in \vec x$, then they neighbor each other as we concluded from \Cref{lemma:general-annotations-dichotomy-case-2}.
    If $y_1, y_2$ are in $\vec x$, they neighbor each other as we showed in the previous paragraph.
    Overall, every two variables in $\neigh(\vec z)$ are neighbors, and from \Cref{lemma:atom-neighbors} we know that there exists an atom in $\Qpfree$ that contains $\neigh(\vec z)$.   
\end{proof}

It is left to show the hardness of \cqstars that are not free-connex or that contain disruptive trios, which easily follows from the hardness of CQs without annotations.

\begin{lem}\label{lem:non-fc-hardness}
    Let $(\K,\oplus,\otimes,\zero,\one)$ be a commutative semiring, and
$Q(\vec x,\star,\vec z)$ a self-join-free \cqstar. If $Q$ is not free-connex or has a disruptive trio, then direct access for $Q$ is not in $\linlog$, assuming the     \HYPERCLIQUE and \sparseBMM hypotheses.
\end{lem}
\begin{proof}
    Consider the CQ $Q'(\vec x,\vec z)$ with the same body as $Q$. According to \Cref{thm:known-dichotomy}, direct access for $Q'$ is not in $\linlog$, assuming the     \HYPERCLIQUE and \sparseBMM hypotheses. We can construct an input for $Q$ given an input for $Q'$ by using the same database and annotating all facts with $\zero$. Due to the semiring properties, we have that $\zero\oplus\zero=\zero$ and $\zero\otimes\zero=\zero$ (since $a\oplus\zero=a$ and $a\otimes\zero=\zero$ for all $a\in\K$), so $Q$ will have the same answers as $Q'$, all annotated by $\zero$, and therefore also in the same order. This way, we can use any direct access algorithm for $Q$ to get direct access for $Q'$ with the same time guarantees. As direct access for $Q'$ is not in $\linlog$, neither is direct access for $Q$.
\end{proof}

Combining \Cref{lemma:general-annotations-order-tractability},  \Cref{lemma:general-annotations-order-dichotomy-full-negative}, and \Cref{lem:non-fc-hardness} completes the proof of \Cref{thm:dichotomy-general-annotation}.

\subsection{Discussion on \ACQs}\label{sec:general:acqs}
Similarly to \Cref{cor:aggregate-tractable}, we conclude from the tractability side of \Cref{thm:dichotomy-general-annotation} the following corollary for \ACQs.
\begin{cor}\label{cor:aggregate-order-tractable}
Let 
$Q(\vec{x}, \alpha(\vec w),\vec{z}) \datarule  \varphi_1(\vec{x},\vec{y}),\dots,\varphi_\ell(\vec{x},\vec{y})$ be a free-connex \ACQ with no disruptive trio. Suppose that $\alpha$ is one of $\aggmin$, $\aggmax$, $\aggcount$, and $\aggsum$.
If $Q$ has an atom that contains $\neigh(\vec z) \cap \free(Q)$, 
then direct access for $Q$ is in $\linlog$.
\end{cor}
Note that, contrasting \Cref{cor:aggregate-tractable}, the aggregation $\aggavg$ is not included in \Cref{cor:aggregate-order-tractable}, since the argument that we can support $\aggavg$ by solving $\aggsum$ and $\aggcount$ separately no longer holds when the aggregation is part of the order. 

In addition, we cannot similarly deduce the hardness of \ACQs based on the negative side of \Cref{thm:dichotomy-general-annotation}.
The proof of \Cref{thm:dichotomy-general-annotation} is based on \Cref{lem:X} that states the hardness of direct access for $R\times S$ by the order of the annotation. For that, we needed to use the power of the annotation, namely, the annotation of an answer $(a,b)$ is the product of the annotations of $R(a)$ and $S(b)$. This does not necessarily imply that we have a similar hardness when the computed value is within an \ACQ, say using $\aggcount$. For example,
direct access for $Q(\aggcount(),x, y) \datarule R(x), S(y)$  is clearly in $\linlog$ since
the computed value has no impact (as it is always $1$).

Nevertheless, we can show that incorporating the computed value in the order introduces hardness for another fixed \ACQ
$\Qaggcount(\aggcount(),x, y)$. Moreover, this \ACQ is tractable if the order was $x,y,\aggcount()$, due to \Cref{thm:annotated-dichotomy}. We prove the hardness of $\Qaggcount(\aggcount(),x, y)$ using a reduction from $\Qstartimes$ (of~\eqref{query:intractable-annotated-order}) with the counting semiring $(\N,+,\cdot,0,1)$. 

\begin{thm}\label{theorem:3SUM-aggregate-hard}
There exists a free-connex \ACQ $\Qaggcount(\aggcount(),x, y)$ such that direct access for $Q$ is not in $\linlog$, assuming the \THREESUM conjecture.
\end{thm}

\begin{proof}
We show a reduction from the \cqstar $\Qstartimes(\star,x,y)$ of Equation~\eqref{query:intractable-annotated-order} over annotated databases $(D,\tau)$ with the semiring
$(\N,+,\cdot,0,1)$. \Cref{lem:X} states that, assuming \THREESUM, $\Qstartimes$ is not in $\linlog$ even under the assumption that, for $N=|D|$, all annotations are within the domain $[1,N^{d+1} - 1]$ for some fixed degree $d$. 
Let $(D,\tau)$ be such an input database for $\Qstartimes$.

We construct an acyclic \ACQ 
$Q_R(x,\aggcount())\datarule \varphi(x,\vec{w_1})$ over a new schema, and show how to construct in linear time a new database $D_R$ for $Q_R$ such that
for every fact $R(a)\in D$ it holds that $(a,n)\in Q_R(D_R)$ if and only if $\tau(R(a))=n$. We similarly define $Q_S(y,\aggcount())\datarule \psi(y,\vec{w_2})$ and $D_S$ for the relation $S$, so that $Q_R$ and $Q_S$ use disjoint sets of relation symbols and variables. Then, the \ACQ $\Qaggcount$ of the theorem is given by
\[\Qaggcount(\aggcount(),x, y) \datarule\varphi(x,\vec{w_1}),\psi(y,\vec{w_2})\,.\]
The database that we construct in the reduction is $D_R\cup D_S$. 
In the remainder of the proof, we show how $Q_R$ and $D_R$ are constructed. The construction of $Q_S$ and $D_S$ is analogous (using a fresh set of variables). 

\begin{figure}
\centering
\begin{tabular}[t]{ccccc}
\multicolumn{5}{l}{$R'$}\\\toprule
$x$ & $u$ & $z_1$ & $z_2$ & $z_3$\\\midrule
2 & 2 & 1 & 1 & 1 \\
13 & 1 & 1 & 1 & 1 \\
13 & 1 & 4 & 1 & 1 \\
13 & 2 & 4 & 1 & 1 \\
64 & 1 & 4 & 4 & 4 \\
192 & 1 & 4 & 4 & 4 \\
192 & 2 & 4 & 4 & 4 \\
\bottomrule
\end{tabular}\quad\quad
\begin{tabular}[t]{cc}
\multicolumn{2}{l}{$T'$}\\\toprule
$u$ & $v$ \\\midrule
1 & 1 \\
2 & 1 \\
2 & 2 \\
\bottomrule
\end{tabular}\quad\quad
\begin{tabular}[t]{cc}
\multicolumn{2}{l}{$T_1$}\\\toprule
$z_1$ & $w_1$ \\\midrule
1 & 1 \\
4 & 1 \\
4 & 2 \\
4 & 3 \\
4 & 4 \\
\bottomrule
\end{tabular}\quad\quad
\begin{tabular}[t]{cc}
\multicolumn{2}{l}{$T_2$}\\\toprule
$z_2$ & $w_2$ \\\midrule
1 & 1 \\
4 & 1 \\
4 & 2 \\
4 & 3 \\
4 & 4 \\
\bottomrule
\end{tabular}\quad\quad
\begin{tabular}[t]{cc}
\multicolumn{2}{l}{$T_3$}\\\toprule
$z_3$ & $w_3$ \\\midrule
1 & 1 \\
4 & 1 \\
4 & 2 \\
4 & 3 \\
4 & 4 \\
\bottomrule
\end{tabular}
\bigskip
$$Q_R(x,\aggcount())\datarule R'(x,u,z_1,\ldots,z_d),T'(u,v),T_1(z_1,w_1),T_2(z_2,w_2),T_3(z_3,w_3)$$
\caption{Example of $Q_R$ and $D_R$ when $N=4$ and $d=3$. The input $(D,\tau)$ is defined by $D = \set{R(2),R(13),R(64),R(192)}$ where $\tau(R(a)) = a$ for every fact $R(a)$.
\label{fig:count}}
\end{figure}

The \ACQ $Q_R$ is the following:
$$Q_R(x,\aggcount())\datarule R'(x,u,z_1,\ldots,z_d),T'(u,v),T_1(z_1,w_1),\ldots,T_d(z_d,w_d)$$
For ease of explanation, we first handle the case of $d=0$.
In this case, the annotations are in $\{1, \ldots, N - 1\}$   and $Q_R$ is $Q_R(x,\aggcount())\datarule R'(x,u),T'(u,v)$.
Consider a fact $R(a)$ of $D$ and its annotation $\tau(R(a))$. As the annotation is in the range $\set{1, \ldots, N-1}$, it can be represented in binary form using $\log N$ bits. 
Let $I\subseteq\set{0, \ldots, \log(N)-1}$ be the set of positions of $1$-bits in the binary representation of the annotation $\tau(R(a))$.
We have that $\tau(R(a))=\sum_{i\in I}{2^i}$.
For every $i\in I$, we put the fact $R'(a, i)$ in $D_R$. We also make sure that there are $2^i$ facts where variable $u$ is assigned value $i$ in $T'$.
More precisely, for every $i\in\ \set{0, \ldots, \log(N)-1}$ and every $j \in \{1, \ldots, 2^i\}$,  we add to $D_R$ the fact $T'(i,j)$.
This construction ensures that the answers to $Q_R$ are exactly the answers to $Q$ where the count is equal to the annotation.

We now define the construction of $Q_R$ and $D_R$ for an arbitrary $d\geq 0$. 
Consider a fact $R(a)$ and its annotation $\tau(R(a))$. As the annotation is in the range $\set{1, \ldots, N^{d+1}-1}$, it can be represented in base $N$ using $d+1$ digits. Each digit in base $N$ can be represented in binary using $\log N$ bits.
Let $p_j$ be the $j$th digit in the base-$N$ representation of $\tau(R(a))$ and let $I_j\subseteq\set{0, \ldots, \log(N)-1}$ be the set of positions of $1$-bits in the binary representation of $p_j$.
For every $i\in I_j$ we add to $D_R$ the following fact:
$$
  R'(a,i,\underbrace{N,\dots,N}_j,\underbrace{1,\dots,1}_{d-j})
$$

For every $i \in 1,\ldots, d$ we make sure there are $N$ tuples starting with $N$ and one tuple starting with $1$ in ${T_i}^{D_R}$.
For that, we add to $D_R$ the fact $T_i(1,1)$, and for every $q=1, \ldots N$ we add the fact $T_i(N,q)$. We create $T'$ as in the case when $d=0$.
Note that the schema of $Q_R$ depends only on $d$. Also note that the facts of $T'$ and $T_i$ depend only on $N$, regardless of the facts of $D$ or their annotations. Finally, the reader can verify that this construction ensures that $(a,n)\in Q_R(D_R)$ if and only if $\tau(R(a))=n$. (See also the following \Cref{exa:construction} for an illustration.)

Finally, let us inspect the size of $D_R$.
Every fact of $D$ is used at most $d\log N$ times in ${R'}^{D_R}$. We also have that $|{T'}^{D_R}|=\sum_{i=0}^{\log N - 1} 2^i = N - 1$, and $|{T_i}^{D_R}|=N+1$. Hence, we add to $D_R$ a total of $O(d \log(N) + N)$ facts.
\end{proof}

The following example illustrates the construction used in the proof of \Cref{theorem:3SUM-aggregate-hard}.

\begin{exa}\label{exa:construction}
\Cref{fig:count} depicts an example of the construction for the database $$D = \set{R(2),R(13),R(64),R(192)}$$ where $\tau(R(a)) = a$ for every fact $R(a)$. In the example, we have $N = |R^D| = 4$ and $d=3$, and thus the construction supports annotations in $[1, \ldots, 4^4 - 1 = 255]$.
We illustrate our construction by showing that for every fact $R(a)$ with the annotation $a$, we will get the answer $(a,a)$ in $Q_R(D_R)$.

\begin{itemize}
\item The fact $R(2)$  has the annotation $2$, which can be represented in binary using a single $1$-bit, so we add to $D_R$ the single fact $R'(2,2,1,1,1)$ and make sure there are two facts in ${T'}^{D_R}$ starting with $2$. So $(2,2)$ will be in $Q_R(D_R)$.
\item For $R(13)$, the annotation $13$ can be represented by the polynomial $3 \cdot 4 + 1$. Therefore, we transform $R(13)$ to three facts in ${R'}^{D_R}$, one with a count of $1 \cdot 1 \cdot 1 \cdot 1$, one with a count of $1 \cdot 4 \cdot 1 \cdot 1$, and one with a count of $2 \cdot 4 \cdot 1 \cdot 1$. As a result, the tuple $(13,13)$ will be in $Q_R(D_R)$ as $1 + 4 + 8 = 13$.
\item As for $R(64)$, its annotation of $64$ can be represented as $1 \cdot 4^3$. As $1$ contains a single active bit, we transform it to the fact $R'(1,4,4,4)$ which has a count of $1 \cdot 4 \cdot 4 \cdot 4 = 64$.
\item Now consider the fact $R(192)$ with the annotation $192$. We can represent $192$ as a polynomial of $N=4$ as $3 \cdot 4^3$ and $3$ can be represented in binary using two $1$-bits. So, the fact $R(192)$ is transformed into the facts $R'(192,1,4,4,4)$ and $R'(192,2,4,4,4)$ in $D_R$. The former of these facts has a count of $1 \cdot 4 \cdot 4 \cdot 4 = 64$ and the latter has a count of $2 \cdot 4 \cdot 4 \cdot 4 = 128$, so the $Q_R(D_R)$ will contain the tuple $(192, 192)$ since $64 + 128 = 192$.
\end{itemize}
Hence, we get the answer $(a,a)$ for each $R(a)$, as we claimed.\qed
\end{exa}

\Cref{lem:X} states hardness of $\Qstartimes(\star,x, y) \datarule R(x),S(y)$, while in \Cref{theorem:3SUM-aggregate-hard} we only claimed the existence of the hard \ACQ $\Qaggcount(\aggcount(),x,y)$, which is more involved than $\Qstartimes(\star,x, y)$ (and we construct it in the proof). The reader might wonder whether we could also phrase \Cref{lem:X} over the Cartesian product $Q(\aggcount(),x,y)\datarule R(x), S(y)$ or alike. Clearly, incorporating $\aggcount$ in
$R(x)\times S(y)$ would be meaningless since every answer appears exactly once (and has a count of $1$). The next theorem shows that the reason goes deeper: even if we add to $Q(\aggcount(),x,y)\datarule R(x), S(y)$ existential variables, the query remains in $\linlog$. 

\begin{prop}\label{prop:annotation-aggregation-hardness-gap}
    For $Q(\aggcount(),x, y) \datarule R(x, w), S(y, z)$, direct access is in $\linlog$.
\end{prop}
\begin{proof}
    For ease of following the proof, we rename $Q$ as follows so that the two atoms have matching structures:
    $$Q(\aggcount(),x, x') \datarule R(x, w), R'(x', w')$$
    In the remainder of this proof, we fix an input database $D$ for $Q$. Note that the answers are of the form $(c\cdot c',a,a')$ where $c$ is the count of the facts $R(a,\cdot)$ that have $a$ as the first element, and $c'$ is the count of the facts $R'(a',\cdot)$ that have $a'$ as the first element. 
\begin{figure}\label{fig:example-cqstar-acq-hardness-gap}
\centering
\renewcommand{\arraystretch}{1.1}
\begin{tabular}[t]{cc}
\multicolumn{2}{l}{$R$}\\\toprule
$x$ & $w$ \\\midrule
$a$ & 1 \\
$b$ & 1 \\
$b$ & 2 \\
$c$ & 1 \\
$c$ & 2 \\
$c$ & 3 \\
\bottomrule
\end{tabular}\quad\quad
\begin{tabular}[t]{cc}
\multicolumn{2}{l}{$R'$}\\\toprule
$x'$ & $w'$ \\\midrule
$a'$ & 1 \\
$b'$ & 1 \\
$c'$ & 1 \\
$c'$ & 2 \\
$d'$ & 1 \\
$d'$ & 2 \\
\bottomrule
\end{tabular}\quad\quad
\parbox{0.5in}{
\centering
\vskip12em
$\longrightarrow$
}
\quad\quad
\begin{tabular}[t]{cccc}
\multicolumn{4}{l}{$L$}\\\toprule
$c$ & $c'$ & $X_c$ & $X'_{c'}$\\\midrule
1 & 1 & $[a]$ & $[a',b']$ \\
1 & 2 & $[a]$ & $[c',d']$ \\
2 & 1 & $[b]$ & $[a',b']$ \\
3 & 1 & $[c]$ & $[a',b']$ \\
2 & 2 & $[b]$ &$[c',d'] $\\
3 & 2 & $[c]$ & $[c',d']$ \\
\bottomrule
\end{tabular}
\bigskip
$$Q(\aggcount(),x, x') \datarule R(x, w), R'(x', w')$$
\caption{Example of the construction in the proof of  \Cref{prop:annotation-aggregation-hardness-gap}: direct access for the \ACQ $Q(\aggcount(),x, x') \datarule R(x, w), R'(x', w')$. \label{fig:tractable_count}}
\end{figure}

    Let us describe the preprocessing step.
    First, we compute the number of facts $R(a,\cdot)$ for every possible value of $a$.
    We use the result to create the set of all counts per possible value of $x$, and denote this set by $C$.
    For all $c\in C$, we also keep a list $X_{c}$ with the set of all values $a$ that appear $c$ times in the left column of $R$. The list $X_c$ is sorted so that we can easily locate a given $a$. We do the same for $R'$ to obtain $C'$ and a sorted list $X'_{c'}$ for every $c'\in C'$.
    
    Next, we create a list $L$, where for every pair of counts $(c,c')\in C \times C'$ it holds the tuple $(c, c', X_{c}, X'_{c'})$ where $X_{c}$ and 
    $X'_{c'}$ are represented as \e{pointers} to the corresponding lists. See \Cref{fig:tractable_count} for an example of the construction of $L$ from an input database.
    We perform direct access on $L$ sorted by $c \cdot c'$ and weighted by $|X_{c}| \cdot |X'_{c'}|$ using prefix sum, similarly to the way established by Carmeli et 
    al.~\cite{carmeli2023tractable} for a single relation, as we briefly describe next.

    We first sort $L$ by the product of the first two elements of each tuple, $c \cdot c'$.
    Notice that $L$ indeed represents the answers in the order we desire. For example, if the first fact in $L$ is $(c, c, X_{c}, X'_{c'})$, then the first $|X_{c}| \cdot |X'_{c'}|$ answers have the count $c \cdot c'$ and assign to $x$ and $x'$ the values that occur in $X_{c}$ and $X'_{c'}$, respectively.
    We then iterate over $L$, and for a tuple index $i$ we compute the sum of $|X_{c}| \cdot |X'_{c'}|$ over all tuples in indices $1, \ldots, i-1$ in $L$. We denote this sum by $l_i$. Note that $l_{i+1}>l_i$ for all $i=1,\dots,|L|$.
    
    We now describe the access procedure. Suppose that we are requested to fetch result number $d$. We perform a binary search on $L$ to find the tuple in index $i$ such that $l_i < d \leq l_{i+1}$. Assume that this tuple is $(c, c', X_{c}, X'_{c'})$. The count we return is $c \cdot c'$. Next, we need to access the $(d-l_i)$th element $(x,x')$ in $X_{c} \times X'_{c'}$, sorted lexicographically by $x$ and then by $x'$. As in multidimensional arrays, we assign to $x'$ the element with the index $(d-l_i)\mod |X'_{c'}|$ of $X'_{c'}$, and we assign to $x$ the element with the index $\lfloor{\frac{d-l_i}{|X'_{c'}|}}\rfloor$ of $X_{c}$.

    We now analyze the execution time.
    Processing each of the $a$ counts and the $a'$ counts separately requires only $O(|D|)$ time. The concern is the time it takes to build and sort the list $L$, which might be of size $|C|\cdot|C'|$. Assume, without loss of generality, that $|C|\geq|C'|$.
    We claim that $|R^D| = \Omega(|C|^2)$.
    If $R^D$ has the smallest possible number of facts to result in $|C|$ distinct counts, then the counts are $1, \ldots,|C|$. The number of facts in this case is $\sum_{i=1}^{|C|}{i}=\frac{|C|(|C|+1)}{2}$.
    So, $|L|=|C|\cdot|C'|\leq |C|^2=O(|R^D|)$.
    We conclude that the algorithm runs in $O(|D| \log |D|)$ preprocessing time and $O(\log |D|)$ access time.  In conclusion, direct access for $Q$ is in $\linlog$, as claimed.
\end{proof}

In the next section, we study how additional assumptions on the annotated database, useful for \ACQs, can lead to additional opportunities to efficiently incorporate the computed value in the ordering.

\section{Locally Annotated Databases}
\label{sec:locally-annotated}

We have seen in \Cref{lem:X} that even the simple \cqstar $\Qstartimes(\star,x, y) \datarule R(x),S(y)$  is intractable, and we have seen in Section~\ref{sec:general:acqs} that this hardness does not necessarily apply to \ACQs. 
As another example, consider the \ACQ
\begin{equation}\label{query:r-annotated-tractable-full-query} 
Q(\aggsum(w),x, y) \datarule R(x,w),S(y)\,.
\end{equation}

When translating into an annotated database, we obtain the \cqstar $\Qstartimes$ over $\Q$-databases annotated by the numerical semiring $(\Q,+,\cdot,0,1)$.  
Hence, we translate the problem into an intractable one. 
Nevertheless, direct access for $Q$ by $(\star, x, y)$ is, in fact, in $\linlog$, as we will show in \Cref{theorem:full-query-annotation-to-variable}. This discrepancy stems from the fact that the hardness of 
$\Qstartimes$ (established in the proof of \Cref{lem:X}) relies on the annotation of tuples from both $R$ and $S$. Yet, in our translation, all $S$-facts are annotated by $1$, and only $R$-facts have a nontrivial annotation. The resulting $\K$-database is such that every fact is annotated by $\one$ (the multiplicative identity), with the exception of one relation. We call such a $\K$-database
 \e{locally annotated} or \e{$R$-annotated} (when we need to specify $R$).  We now focus on such databases and show how the assumption of local annotations can be used for efficient access. 

In the remainder of this section, we restrict the discussion to queries without self-joins\footnote{In fact, for the algorithm, we only need that at most one atom refers to a relation with unrestricted annotations, so it suffices that the relation with unrestricted annotations is not part of more than one atom. The lower bounds do rely on the assumption that every refers to a distinct relation, just like the lower bound of \Cref{thm:known-dichotomy} and all the other lower bounds in this paper.} and
fix a logarithmic-time commutative semiring $(\K,\oplus,\otimes,\zero,\one)$. 

\subsection{Full \cqstars.}
We first discuss full \cqstars (i.e., without existential variables). In the next sections, we also discuss the implications on (non-full) \ACQs.  We begin with some examples that demonstrate the results that follow later in this section.

\begin{exa}\label{example:makes-intractable}
In some cases, incorporating the annotation in an otherwise tractable order may introduce hardness.
Let $(\K,\oplus,\otimes,\zero,\one)$ be a logarithmic-time commutative semiring, and consider the full \cqstar
$$Q(\vec{x}, \star,\vec{z}) \datarule R(v_1, v_3),S(v_2, v_3)$$
over $S$-annotated $\K$-databases. Note that $Q$ has a disruptive trio if $v_3$ appears after both $v_1$ and $v_2$ in the order.
Let us consider orders where this is not the case.
The lexicographic order $(\star,v_2,v_3,v_1)$ is not covered by \Cref{thm:dichotomy-general-annotation}, as $v_1$ appears only in $R$, but $v_2$ does not. However, we will show in \Cref{theorem:full-query-annotation-to-variable} that, when considering $S$-annotated $\K$-databases, direct access for $Q$ by this order is in $\linlog$, while direct access for $Q$ by $(\star,v_1,v_3,v_2)$ is \e{not} in $\linlog$.
\qed\end{exa}

\begin{exa}\label{example:r-annotated-tractable-full-query-fd} 
It may also happen that we incorporate the annotation non-trivially and the query remains tractable. 
Consider the full \cqstar $$Q'(\vec{x}, \star,\vec{z}) \datarule R(v_1, v_3),S(v_2, v_3),T(v_3)$$ over $T$-annotated $\K$-databases. Note that this case is a slight variation on \Cref{example:makes-intractable}. For any lexicographic order $(\vec{x}, \star,\vec{z})$, if $v_3$ appears after $v_1$ and $v_2$, then $Q'$ has a disruptive trio and, therefore, direct access for $Q'$ is not in $\linlog$. As we show in \Cref{theorem:full-query-annotation-to-variable}, for $Q'$ over $T$-annotated $\K$-databases, the lack of a disruptive trio is a sufficient condition for tractability. That is, if $v_3$ does not appear after $v_1$ and $v_2$, then it holds that direct access for $Q'$ (and for $Q$ of \Cref{example:makes-intractable}) is in $\linlog$.
\qed\end{exa}

When dealing with $R$-annotated databases, we can replace the annotation of the facts of $R$ with a new extra attribute, added to $R$, and then reason about orders that involve the annotation by considering orders that involve the new attribute instead. To do so, we introduce the following variations of a query and order.
Let $Q$ be a full acyclic \cqstar without self-joins. For a relation symbol $R$ of $Q$, we define the \e{$R$-deannotation} of $Q(\vec x, \star, \vec z)$ to be the CQ $Q_R$ obtained as follows, where we denote by $\varphi_S$ the atom of a relation symbol $S$.
\begin{itemize}
    \item In the head, replace $\star$ with a new variable $y$.
    \item If, in addition, $\var(\varphi_R)$ contains only variables from $\vec x$, then in the head of $Q_R$, advance $y$ to be immediately after the last variable of $\varphi_R$.
    \item For each relation $S$ of $Q$, if $\var(\varphi_R) \subseteq \var(\varphi_S)$ then concatenate $y$ to the variable sequence of $\varphi_S$.
\end{itemize}

\begin{exa}\label{example:r-deannotation-qprime}
Consider the \cqstar $Q'$ of \Cref{example:r-annotated-tractable-full-query-fd}. The $R$-deannotation of $Q'$ is
$$Q_R(\vec{x}_R, y,\vec{z}_R) \datarule U(v_1, v_3,y),V(v_2, v_3,y),R(v_3,y)$$
where $\vec{x}_R$ and $\vec{z}_R$ are adjustments of $\vec{x}$ and $\vec{z}$: if $v_3$ is in $\vec{x}$, the suffix of $\vec{x}$ that follows $v_3$ is moved to the beginning of $\vec{z}$.
\qed\end{exa}

When $Q$ is full, we can reduce direct access for $Q$ to 
direct access for $Q_R$. This is stated in the next theorem. The theorem also says that, whenever the annotation domain contains the natural numbers, this reduction is optimal in the sense that, if we got an intractable $Q_R$, then direct access for $Q$ was hard to begin with.

\begin{thm}\label{theorem:full-query-annotation-to-variable}
Let $(\K,\oplus,\otimes,\zero,\one)$ be a logarithmic-time commutative semiring.
Let $Q$ be a full \cqstar without self-joins and
$Q_R$ the $R$-deannotation of $Q$ for a relation symbol $R$ of $Q$. 
\begin{enumerate}
\item If $Q_R$ is acyclic and has no disruptive trio, then direct access for $Q$ is in $\linlog$ on $R$-annotated $\K$-databases.\label{theorem:full-query-annotation-to-variable-item1}
\item Otherwise, if $\N \subseteq \K$, then direct access for $Q$ is not in $\linlog$ on $R$-annotated $\K$-databases, assuming the
    \HYPERCLIQUE and \sparseBMM hypotheses.\label{theorem:full-query-annotation-to-variable-item2}
\end{enumerate}
\end{thm}

We note that the negative side of \Cref{theorem:full-query-annotation-to-variable} applies to any domain other than $\N$, as long as we can generate infinitely many elements according to the underlying order of the semiring. Before we prove the theorem, let us give a usage example.

\begin{exa}\label{example:r-deannotation}
\Cref{theorem:full-query-annotation-to-variable} gives us a useful tool to analyze the previous examples. 
Recall that in \Cref{example:r-deannotation-qprime} we showed the $R$-deannotation of $Q'$ from \Cref{example:r-annotated-tractable-full-query-fd}.
Since $y$ appears in every atom, it cannot be part of any disruptive trio. In particular, the disruptive trios of $Q_R$ are exactly the disruptive trios of $Q$. Therefore, given an order 
$(\vec{x}, \star ,\vec{z})$, checking whether direct access for $Q$ is in $\linlog$ boils down to verifying that $v_1$, $v_2$ and $v_3$ do not form a disruptive trio.
\qed \end{exa}

To prove \Cref{theorem:full-query-annotation-to-variable}, 
we first prove the following lemma.
\begin{lem}\label{lemma:deannotation-acyclic}
    Let $(\K,\oplus,\otimes,\zero,\one)$ be a logarithmic-time commutative semiring.
    Let $Q$ be a full \cqstar without self-joins and
    $Q_R$ the $R$-deannotation of $Q$ for a relation symbol $R$ of $Q$. 
    If $Q$ is acyclic, then $Q_R$ is acyclic.
\end{lem}

\begin{proof}
    Since $Q$ is acyclic it has a join tree $T$. We add the variable $y$ to every vertex of $T$ that contains all variables of $R$. As per the definition of $R$-deannotation each vertex in the tree now corresponds to an atom of $Q_R$. We know that every variable in $T$, besides $y$, satisfies the running-intersection property. This is true, in particular, for every variable in $R$. Since the intersection of subtrees is in itself a subtree, we get that $y$ is added to a connected set of vertices, and it too satisfies the running-intersection property. Hence, the new tree is a join tree of $Q_R$, and so, $Q_R$ is acyclic.
\end{proof}

With \Cref{lemma:deannotation-acyclic} at hand, we can now prove \Cref{theorem:full-query-annotation-to-variable}
\begin{proof}
    We prove each part separately. 
    \subparagraph*{Part 1.}
    First, we show that if $Q_R$ is acyclic and has no disruptive trio, then direct access for $Q$ on $R$-annotated $\K$-databases is in $\linlog$.
    Let $(D, \tau)$ be an $R$-annotated $\K$-database. Since $Q$ is full, the answer $Q(D, \tau)$ is simply the join of the relations of all atoms in $Q$. Since $D$ is also $R$-annotated, the annotation of an answer is the annotation of its supporting fact in $R^D$. Yet, the annotation of an $R$-fact is functionally dependent on the other attributes. We will use a recent result on direct access in the presence of Functional Dependencies (FDs)~\cite{carmeli2023tractable}.
    
    We first construct from the \cqstar $Q$ and the annotated database $(D, \tau)$ a CQ $Q\fd$ and a database $D\fd$, as follows:
    \begin{itemize}
        \item Introduce a relation symbol $R_e$ of arity $\ar(R)+1$. 
        \item Modify the atom containing $R$ in $Q$ by replacing $R$ with $R_e$ and concatenating a new variable $y$.
        \item Construct $R_e^{D\fd}$ from $R^{D}$ by setting the value of $y$ in each fact $f$ to be $\tau(f)$.
        \item In the head, replace $\star$ with $y$.
         \item Add the FD $R_e: 1,\ldots,\ar(R) \rightarrow \ar(R)+1$ to the schema of $Q\fd$. (Hence, here we allow the schema to include an integrity constraint.) 
    \end{itemize}   
    From the discussion above, answers in $Q\fd(D\fd)$ are identical to answers in $Q(D, \tau)$, except for the annotations that are now represented as values for $y$. Hence, if we prove that direct access for $Q\fd$ is in $\linlog$, then so is direct access for $Q$. We do so next.
    
    We follow the algorithm suggested by Carmeli et al.\cite{carmeli2023tractable} to obtain what they refer to as the \e{FD-reordered extension} of $Q\fd$. By doing so, we get \e{precisely} our $R$-deannotation $Q_R$ of $Q$. Carmeli et al.~\cite{carmeli2023tractable} proved that given a CQ and its FD-reordered extension, if direct access for the FD-reordered extension is in $\linlog$, then so is direct access for the original CQ, which is $Q\fd$ in our case. Therefore if direct access for $Q_R$ is in $\linlog$, then so is
    direct access for $Q\fd$, and so is direct access for $Q$ on $R$-annotated $\K$-databases. Since $Q_R$ is acyclic and has no disruptive trio, by \Cref{thm:known-dichotomy} direct access for $Q_R$ is, indeed, in $\linlog$.

    \subparagraph*{Part 2.}
    Next, we assume that $\N \subseteq \K$ and either $Q_R$ is cyclic or $Q_R$ has a disruptive trio.
    We analyze all possible cases and show that in every case direct access for $Q$ on $R$-annotated $\K$-databases is not in $\linlog$ (under the corresponding complexity assumptions).
    We denote by $X_R$ the set of variables that occur in the atom of $R$ in $Q$. 

    \parcase{Case 1}{$Q_R$ is cyclic.}
    In \Cref{lemma:deannotation-acyclic} we proved that if $Q$ is acyclic, so is $Q_R$. Therefore, if $Q_R$ is cyclic, so is $Q$, and by \Cref{thm:annotated-dichotomy} direct access for $Q$ on $R$-annotated $\K$-databases is not in $\linlog$, assuming the \HYPERCLIQUE hypothesis.
    In the remaining cases, we assume that $Q_R$ is acyclic.

    \parcase{Case 2}{There exists a disruptive trio that does not involve $y$.}
    Next, we assume that $x_1$, $x_2$, and $x_3$ form a disruptive trio in $Q_R$, and $y$ is not one of $x_1,x_2,x_3$. We create a new CQ $Q_0$ by removing $y$ from $Q_R$. The variables $x_1,x_2,x_3$ still form a disruptive trio in $Q_0$, and we know by \Cref{thm:known-dichotomy} that direct access for $Q_0$ is not in $\linlog$, assuming \sparseBMM. Assume, by way of contradiction, that direct access for $Q$ on $R$-annotated $\K$-databases is in $\linlog$. The lexicographic order of the \cqstar $Q$ differs from that of the CQ $Q_0$ by the existence of $\star$; other than that, $Q_0$ is identical to $Q$. Therefore, we can provide direct access for $Q_0$ on a given database $D$ by annotating each fact in $D$ with $\one$, and providing direct access for $Q$ on the annotated database. We conclude that direct access for $Q_0$ is in $\linlog$, which is a contradiction. 
    
    \parcase{Case 3}{There is a disruptive trio that involves $y$, and $y$ is a neighbor of both other trio members.}
    Suppose that $x_1$, $x_2$, and $y$ form such a disruptive trio in $Q_R$. Then  $y$ is a neighbor of both $x_1$ and $x_2$, the variables $x_1$ and $x_2$ are non-neighbors, and $y$ appears after $x_1$ and $x_2$ in the order. Recall that $X_R$ denotes the set of variables in the atom of $R$. Denote by $x_f$ the variable in $X_R$ that appears last in $Q_R$.
    We claim that $x_1$, $x_2$, and $x_f$ form a disruptive trio in $Q_R$.
    Since $y$ was added only to atoms containing all variables in $X_R$, we have that $x_f$ appears in all atoms containing $y$, and so, both $x_1$ and $x_2$ are neighbors of $x_f$. Since $x_1$ and $x_2$ are not neighbors of each other, we have that $x_1 \neq x_f$ and $x_2 \neq x_f$. 
    We know that $x_f$ appears either after $y$ or immediately before $y$ in the head since this is how the head of $Q_R$ was reordered. Since $x_f$ appears in $Q_R$ either after $y$ or immediately before $y$, it necessarily appears after $x_1$ and $x_2$. This proves that $x_1$, $x_2$, and $x_f$ form a disruptive trio in $Q_R$. As we saw in the previous case, a disruptive trio that does not include $y$ means direct access for $Q$ on $R$-annotated $\K$-databases is not in $\linlog$, assuming \sparseBMM.

    \parcase{Case 4}{There is a disruptive trio that involves $y$, and $y$ is a neighbor of only one of the other trio members.}
    Assume that $x_2$ is the neighbor of both $x_1$ and $y$, that $x_1$ and $y$ are not neighbors, and that $x_2$ appears after $x_1$ and $y$ in the order. 
    
    We claim that since $x_1$ and $y$ are not neighbors, there exists a variable $x' \in X_R$ that is not a neighbor of $x_1$ in $Q_R$. 
    Assume, by way of contradiction, that no such variable exists and so every variable of $X_R$ is a neighbor of $x_1$. By \Cref{lemma:atom-neighbors}, there exists an atom in $Q_R$, and by extension $Q$ that contains $\set{x_1} \cup X_R$. Recall that we constructed $Q_R$ as the $R$-deannotation of $Q$ by adding $y$ to every atom that contains $X_R$. Therefore, $x_1$ and $y$ occur jointly in the atom that contains  $\set{x_1} \cup X_R$ and are neighbors.
    We conclude that since $x_1$ and $y$ are not neighbors, there exists a variable $x' \in X_R$ such that $x_1$ and $x'$ are not neighbors. 
    
    As $y$ appears only in atoms that contain $X_R$, and $x'$ is a variable in $X_R$, we have that $x'$ is in every atom that contains $y$. Therefore, since $x_2$ is a neighbor of $y$, it is also a neighbor of $x'$. Overall, we have that $x_2$ is a neighbor of both $x'$ and $x_1$, but $x_1$ is not a neighbor of $x'$.
    Assume first that $x'$ appears before $y$ in the head of $Q_R$. As $y$ appears before $x_2$, we have that $x'$ appears before $x_2$.
    Therefore, the variables $x_1$, $x'$, and $x_2$ form a disruptive trio with respect to $Q_R$. We know that if there exists a disruptive trio that does not involve $y$, then direct access for $Q$ by on $R$-annotated $\K$-databases is not in $\linlog$, assuming \sparseBMM.
    So now, we are left with the case where $x'$ appears after $y$. In the remainder of the proof, we handle this case.
    
    We create a new CQ $Q_0$ from $Q_R$ by removing $y$ and moving $x'$ to the position that $y$ occupied in the head. As $x_2$ appears in $Q_R$ after $x_1$ and $y$, it appears in $Q_0$ after $x_1$ and $x'$. Since $x_2$ is a neighbor to both $x_1$ and $x'$, and $x'$ and $x_1$ are not neighbors, it holds that $x_1$, $x'$, and $x_2$ form a disruptive trio in $Q_0$. From \Cref{thm:known-dichotomy}, we know that direct access for $Q_0$ is not in $\linlog$, assuming \sparseBMM. 
    To conclude the proof, we show a reduction from direct access for $Q_0$ to direct access for $Q$ on $R$-annotated $\K$-databases. This would immediately imply that direct access for $Q$ on $R$-annotated $\K$-databases is not in $\linlog$, assuming \sparseBMM. 

    Recall our assumption that the semiring domain includes $\N$. We take all values of $x'$ in $R^D$, place them in an array, denoted by $A$, and sort them. We denote by $h(f)$ the position of $f$ in $A$. Then we add annotations to $D$. For every fact $f$ in $R^D$, we define that $\tau(f) = h(f)$. For every fact $f$ in $D$ but not in $R^D$, we define $\tau(f) = \one$.
    Note that, with this construction, ordering by $\star$ is the same as ordering by $x'$. As $x'$ is in $X_R$ and appears after $\star$ in the order, we have that $y$ is not advanced during the construction of $Q_R$, and so, $\star$ appears in $Q$ in the same position where $y$ appears in $Q_R$. Since $x'$ appears after $y$ in $Q_R$, it also appears after $\star$ in $Q$. Due to the way we annotate the facts, sorting the results first by $\star$ and then by $x'$ is the same as sorting only by $\star$.
    This means that, for all $i$, the $i$-th answer for $Q_0(D)$ is the same as the $i$-th answer for $Q(D, \tau)$, where the value of the annotation in $Q(D, \tau)$ is the position in $A$ of the value assigned to $x'$ in $Q_0(D)$, and the value for $x'$ appears in the answer of $Q(D, \tau)$ in a different position.
    When the $i$-th answer for $Q_0(D)$ is requested, we fetch the $i$-th answer for $Q(D, \tau)$ and
    replace the annotation with the value in the position of $x'$.  Hence, we get the claimed reduction from direct access for $Q_0$ to direct access for $Q$ on $R$-annotated $\K$-databases, where the database construction takes $O(|D|\log|D|)$ time, and the translation of an answer during access takes constant time.
\end{proof}

\subsection{Queries with projection in the case of idempotence.}\label{subsection:general-queries-idempotence}
Next, we extend \Cref{theorem:full-query-annotation-to-variable} beyond full \cqstars. 
In this section, we will be working with queries without self-joins. For simplicity, we will often mention a relation symbol to refer to the atom using this relation symbol. 
Consider the following \ACQ:
\begin{equation}\label{query:r-annotated-tractable-query} 
Q(\aggmax(w_2), x_1, x_2, x_3) \datarule R(x_1, x_3,w_3),S(x_2, x_3),T(x_3, w_1),U(w_1,w_2)
\end{equation}
We can solve direct access for $Q$ using direct access for
the \cqstar
$$Q'(\star, x_1,  x_2, x_3) \datarule R(x_1, x_3,w_3),S(x_2, x_3),T(x_3, w_1),U'(w_1)$$ over $U'$-annotated $\Q$-databases and the max tropical semiring.
We can then eliminate existential variables as we did in \Cref{lemma:existential-elimination-semiring-Qfree} and reduce direct access for $Q'$ to direct access for the full \cqstar $\Qtpfree$. 
$$\Qtpfree(\star, x_1, x_2, x_3) \datarule \pfreeQp{R}(x_1, x_3),\pfreeQp{S}(x_2, x_3),\pfreeQp{T}(x_3)$$
As we will show in \Cref{lemma:existential-removal-idempotent-positive}, when the input database for $Q'$ is $U'$-annotated over the max tropical semiring, the suitable database for $\Qtpfree$ is $\pfreeQp{T}$-annotated.
From $\Qtpfree$ we define $Q_0$ as the $\pfreeQp{T}$-deannotation of $\Qtpfree$.
$$Q_0(y, x_1, x_2, x_3) \datarule R_0(x_1, x_3, y),S_0(x_2, x_3, y),T_0(x_3, y)$$
Since $\Qtpfree$ is full and $Q_0$ is acyclic with no disruptive trios, \Cref{theorem:full-query-annotation-to-variable} implies that $\Qtpfree$ is in $\linlog$ on $\pfreeQp{T}$-annotated $\Q$-databases. We therefore get that direct access for $Q'$ on $U$-annotated $\Q$-databases is in $\linlog$, and as a consequence so is direct access for $Q$.

As we explain next, the argument above is specific to $\aggmax$ and not all aggregate functions since we rely on $\aggmax$
being \e{idempotent}. 
An operation $\oplus$ is said to be idempotent if for every $a$ in the domain $\K$ we have that $a \oplus a = a$.
We say that a commutative semiring is an \e{$\oplus$-idempotent semiring} if its addition operation, $\oplus$, is idempotent.

Let $(\K,\oplus,\otimes,\zero,\one)$ be some logarithmic-time commutative semiring.
The process of existential variable elimination as given in \Cref{lemma:existential-elimination-semiring-Qfree} takes a free-connex \cqstar $Q$ without self-joins and a $\K$-database $(D, \tau)$ and translates it to a full acyclic \cqstar $\Qpfree$ and a $\K$-database $(D_\free, \tau_\free)$.
When working over an $\oplus$-idempotent semiring, if the input database is locally annotated, then we can show that the output database is also guaranteed to be locally annotated. The semirings used for $\aggcountd$ (over logarithmic domain), $\aggmin$, and $\aggmax$ are $\oplus$-idempotent and therefore provide such a guarantee, while the semirings used for $\aggsum$ and $\aggcount$ do not. In the case of the query of~\eqref{query:r-annotated-tractable-query}, 
if the aggregate function was $\aggsum$ instead of $\aggmax$, once we eliminate existential variables, the database for $\Qpfree$ is no longer guaranteed to be locally-annotated since projecting out the existential variable $w_3$ may cause $R$, in addition to $T'$, to be annotated with values other than $\one$. 

Let $Q$ be a free-connex \cqstar without self-joins using a relation symbol $R$. We define the \emph{carrying variables} of $R$ to be the free variables $x$ such that there is a path in $H(Q)$ from $x$ to a variable of $R$ where all variables on the path other than $x$ are existential. Notice that all free variables of the atom using $R$ are carrying, as witnessed by a path of length $0$. Notice also that there might be no carrying variables in case the query is Boolean or in case $R$ contains only existential variables and the query hypergraph is disconnected (e.g., $Q(y) \datarule R(x),S(y)$). 

\begin{lem}\label{lem:carrier-exists}
Let $Q$ be a free-connex \cqstar without self-joins and using a relation symbol $R$. There exists an atom in $Q$ whose free variables are precisely the carrying variables of $R$.
\end{lem}
\begin{proof}
    Let us first consider the case that the $R$ atom contains only free variables. In this case, the carrying variables are exactly the variables of $R$, so the $R$ atom satisfies the lemma requirements, and we are done. 
    Assume next that $R$ contains existential variables.
    
    Since $Q$ is free-connex, there is a join tree $T$ containing a vertex for each of its atoms and another vertex $v$ containing exactly $\free(Q)$. Consider the path $P$ in $T$ from $v$ to the vertex of $R$. We claim that the first vertex on $P$ containing an existential variable that is connected by a path of only existential variables to a variable of $R$ satisfies the lemma.\footnote{This definition may seem overly complicated, but it is there to handle cases like the query $Q(x_1,x_2) \datarule S(x_1,x_2,y_1),R(x_1,y_2)$ and the extended join tree $(x_1,x_2)-(x_1,x_2,y_1)-(x_1,y_2)$. Here, only $x_1$ is carrying.} We denote this vertex by $u$.
    Such a vertex exists since we assume that $R$ contains existential variables, and so the vertex of $R$ has such a path of length $0$.

    We can easily see that all free variables in $u$ are carrying. Indeed, let $x$ be a free variable of $u$. Since $u$ contains an existential variable that has a path of only existential variables to a variable of $R$, concatenating $x$ to the beginning of this path is a witness to the fact that $x$ is carrying.

    It is left to prove that all carrying variables appear in $u$.
    We treat $T$ as rooted in $v$, and denote by $T_u$ its subtree rooted in $u$. Notice that the $R$ atom is in $T_u$.
    Let $w_1$ be a carrying variable, and consider a path $w_1,\ldots,w_k$ witnessing this (i.e., $w_i$ is existential for all $i>1$, and $w_k$ is in the $R$ atom). We claim by induction on decreasing $i$ that every variable $w_i$ on this path appears in $T_u$.
    First, $w_k$ appears in $T_u$ as it appears in the $R$ atom. Next, consider $w_i$ with $i>1$ that appears in $T_u$. The variable $w_i$ cannot appear in the parent of $u$ because $u$ is the first vertex on $P$ that is allowed to contain a variable starting a path of existential variables such as $w_i,\ldots,w_k$. Since $w_i$ appears in $T_u$ but not in the parent of $u$, by the running intersection property, we have that all atoms containing $w_i$ appear in $T_u$. This includes the atom witnessing the fact that $w_i$ and $w_{i-1}$ are neighbors. Thus, we have that $w_{i-1}$ also appears in $T_u$. From this induction, we deduce that $w_1$ appears in $T_u$. Since it appears in $T_u$ and in $v$, by the running intersection property, it must appear in $u$.
\end{proof}

\Cref{lem:carrier-exists} shows that there is an atom of $\Qpfree$ that contains exactly the carrying variables. We denote the relation symbol of this atom by $R\carry$.
Next, we prove the positive side of the dichotomy using a construction similar to \Cref{lemma:existential-elimination-semiring-Qfree} but specific to locally annotated databases with $\oplus$-idempotent semirings.

\begin{lem}\label{lemma:existential-removal-idempotent-positive}
Let $(\K,\oplus,\otimes,\zero,\one)$ be a logarithmic-time $\oplus$-idempotent commutative semiring, and let $R$ be a relation symbol of a free-connex \cqstar $Q$ without self-joins. If direct access for $\Qpfree$ over $R\carry$-annotated $\K$-databases is in $\angs{T_p,T_a}$, then direct access for $Q$ over $R$-annotated $\K$-databases is in $\angs{\loglin+T_p,T_a}$.
\end{lem}

To prove \Cref{lemma:existential-removal-idempotent-positive}, we notice that when our construction from \Cref{lemma:existential-elimination-semiring-Qfree} is applied to a locally annotated database with $\oplus$-idempotent semirings, the result is a locally annotated database, where the annotation is on an atom that contains all carrying variables and depends only on them. We can then move the annotations to guarantee that the construction results in an $R\carry$-annotated database.

\begin{proof}
We show an $O(|D|\log|D|)$-time construction that maps $R$-annotated $\K$-databases $(D, \tau)$ of $Q$ to $R\carry$-annotated $\K$-databases $(D', \tau')$ of $\Qpfree$ such that $\Qpfree(D', \tau') = Q(D, \tau)$.
    Given an $R$-annotated $\K$-databases of $Q$, we use our construction from \Cref{lemma:existential-elimination-semiring-Qfree}.
    Recall that as a first step, we assign a relation to every vertex of an ext-free-connex tree. These relations are annotated either by the original annotations or by $\one$, so at this point, only the one vertex matching $R$ in the tree is annotated by non-trivial annotations. Next, until we are left with only free variables, we repeatedly eliminate some leaf vertices as follows:
\begin{enumerate}
    \item In the relation associated with the vertex, project out every variable not shared between the vertex and its neighbor while summing ($\oplus$) annotations.
    \item Join the obtained relation into the relation of the neighbor vertex while multiplying ($\otimes$) annotations.
    \item Remove the leaf from the tree, its associated atom from the query, and its associated relation from the database.
\end{enumerate}
    
    We first claim that our construction results in a locally-annotated database. If we eliminate a relation that is only annotated by $\one$, then since the semiring is $\oplus$-idempotent (and in particular $\one\oplus\one=\one$), we get that its projection is still only annotated by $\one$, and when we join the projection into another relation, if it was only annotated by $\one$, it remains this way.
    When we eliminate a relation with unrestricted annotations, we know that it is the only relation annotated by values other than $\one$; thus, at the end of the step, the database is still locally annotated, except now the relation with unrestricted annotation is its parent in the tree.

    We next claim that the obtained annotation does not depend on non-carrying free variables. In fact, we prove a slightly stronger claim. We say that a variable is relevant if it is either carrying or it is an existential variable that has a path to an existential variable of $R$ that only goes through existential variables. Notice that any variables that appear in the same atom as an existential relevant variable are also relevant. We claim that, during the elimination process, whenever a relation contains two facts that differ only on non-relevant variables, these two facts have the same annotation. 
    Since the elimination process results in only free variables, we get that at its end, the annotation depends only on carrying variables.
    
    This claim is trivially true in the beginning since all variables in $R$ are relevant and all other relations are always annotated by $\one$.
    Consider the step of eliminating a vertex $v_e$ into its parent $v_p$, and assume as an induction hypothesis that the claim holds before this step. We want to show that the claim also holds in $v_p$ after this step (this is enough because $v_e$ is eliminated in this step and all other relations remain unchanged). We consider two cases.
    \begin{enumerate}
        \item 
        The first case is that all variables in $v_e$  are relevant. In particular, all shared variables between $v_p$ and $v_e$ are relevant. Thus, when joining the projected $v_e$ into $v_p$, facts in $v_p$ that disagree only on non-relevant variables, agree on the shared variables with $v_e$, will be multiplied by the same annotation, and will keep having identical annotations to one another.
    \item
    The second case is that $v_e$ contains a non-relevant variable. From the induction hypothesis, before the projection of $v_e$, the annotations in $v_e$ do not depend on the shared non-relevant variables. Since $v_e$ contains a non-relevant variable, we have that $v_e$ does not contain a relevant existential variable, or in other words, all existential variables in $v_e$ are non-relevant.
    From the induction hypothesis, before the projection of $v_e$, the annotations in $v_e$ do not depend on any existential variables. Due to the idempotence, the projection does not change the annotation of the resulting facts.
    Thus, after the projection, the annotation in $v_e$ still does not depend on the shared non-relevant variables.
    When joining the projected $v_e$ into $v_p$, facts in $v_p$ that disagree only on non-relevant variables would be multiplied by the same annotations, and so will keep having identical annotations to one another.
\end{enumerate}

    Since the nontrivial annotations appear in only one atom and may only depend on carrying variables, we can move them to any atom containing all carrying variables, and in particular $R\carry$.
\end{proof}

\Cref{lemma:existential-removal-idempotent-positive} proves the positive direction. That is, assuming a $\oplus$-idempotent commutative semiring, if we want to know that $Q$ over $R$-annotated databases admits efficient direct access, then it is enough to require that $\Qpfree$ over $R\carry$-annotated databases admits efficient direct access. We next prove the other direction.

\begin{lem}\label{lemma:existential-removal-idempotent-negative}
Let $(\K,\oplus,\otimes,\zero,\one)$ be a commutative semiring, and let $R$ be a relation symbol of a free-connex \cqstar $Q$ without self-joins.
If direct access for $Q$ over $R$-annotated $\K$-databases is in $\angs{T_p,T_a}$, then direct access for $\Qpfree$ over $R\carry$-annotated $\K$-databases is in $\angs{\lin+T_p,T_a}$.
\end{lem}
We first notice that we cannot simply use the same construction as in the proof of \Cref{lem:reduction} to prove \Cref{lemma:existential-removal-idempotent-negative}. There, we assigned a value to all existential variables, so we do not necessarily have that answers with distinct annotations use distinct facts of $R$, and we cannot simply move the annotations from $R\carry$ to $R$. Instead, we use a different construction that propagates the values assigned to the variables of $R\carry$ through the existential variables into the variables of $R$, thus allowing us to move the annotations from $R\carry$ to $R$.

\begin{proof}
We show an $O(|D'|)$-time construction that maps $R\carry$-annotated $\K$-databases $(D', \tau')$ of $\Qpfree$ to $R$-annotated $\K$-databases $(D, \tau)$ of $Q$ such that $\Qpfree(D', \tau') = Q(D, \tau)$.
As a first step, we perform the Yannakakis full reduction~\cite{DBLP:conf/vldb/Yannakakis81}, and then we can assume that every fact $f'$ of $D'$ is consistent with $\Qpfree$ in the sense that we can obtain $f'$ from the corresponding atom of $\Qpfree$ by assigning some values to the variables. Then, we can denote by $f'[x]$ the value of $f'$ in the positions of the variable $x$.
We also denote by $\langle f' \rangle$ a value that encodes the tuple of the fact $f'$.

Recall the definition of carrying variables. The atom of $R\carry$ contains all free variables of $Q$ that have a path in $H(Q)$ to a variable of the atom of $R$ where every variable on the path, aside from the carrying variable, is existential. Denote by $V$ the set of all existential variables that are part of such a path.
Note that if an atom $\varphi$ of $Q$ contains a variable in $V$, then every free variable in $\varphi$ must be a carrying variable by definition, so the free variables of $\varphi$ all appear in the atom of $R\carry$.
Recall that $\Qpfree$ is created directly from $Q$ by restricting $Q$ to the free variables, so every atom $\varphi$ in $Q$ has a matching atom $\pfreeQ{\varphi}$ in $\Qpfree$.
Let us now construct $(D, \tau)$ as follows.

\bigskip
\begin{algorithmic}[1]
\hrule
\STATE Remove dangling facts using the Yannakakis algorithm~\cite{DBLP:conf/vldb/Yannakakis81}

\FORALL{atoms $\varphi$ of $Q$}\label{line:iterate-q}
    \IF{$\varphi$ has a variable in $V$}\label{line:choose-data-source}
        \STATE $\mathrm{source} \defeq R\carry$
    \ELSE
        \STATE $\mathrm{source} \defeq$ the relation symbol of $\pfreeQ{\varphi}$ in $\Qpfree$
    \ENDIF
    \FORALL{facts $f'$ in the relation of $\mathrm{source}$ in $D'$}\label{line:iterate-over-pfree}
        \FORALL{$x\in\var(\varphi)$}\label{line:iterate-over-atom-variables}
            \IF{$x\in\free(\varphi)$}
                \STATE $c_x\defeq f'[x]$\label{line:qpfree-free-variables-agreement}
            \ELSIF{$x \in V$}
                \STATE $c_x \defeq \langle f' \rangle$\label{line:qpfree-excarry-variables}
            \ELSE
                \STATE $c_x \defeq a_0$\label{line:qpfree-existential-variables}
            \ENDIF
        \ENDFOR
        \STATE Add to $D$ a new fact $f$ obtained from $\varphi$ by replacing each variable $x$ with $c_x$\label{line:add-agree-fact-carry}
        \IF{the relation symbol of $\varphi$ is $R$}
            \STATE $\tau(f) \defeq$ the annotation of the fact agreeing with $f'$ in $R\carry$\label{line:carry-annotation}
        \ELSE
            \STATE $\tau(f) \defeq \one$\label{line:non-carry-annotations}
        \ENDIF
    \ENDFOR
\ENDFOR
\hrule
\end{algorithmic}
\bigskip

To create $D$, we iterate over every atom of $Q$ (Line~\ref{line:iterate-q}) and construct its relation.
We begin by choosing the source of the data for the relation (Line~\ref{line:choose-data-source}). If $\varphi$ contains a variable in $V$, we set its data source to be $R\carry$. Otherwise, we set its data source to be $\pfreeQ{\varphi}$.
Then, we iterate over the facts of the data source in $D'$ (Line~\ref{line:iterate-over-pfree}), and for each fact $f'$, we create in the relation of $\varphi$ in $D$ a new fact $f$. The new fact $f$ is created so that it agrees on the free variables with $f'$ (Line~\ref{line:qpfree-free-variables-agreement}), has an encoding of $\langle f' \rangle$ in place of the variables of $V$ (Line~\ref{line:qpfree-excarry-variables}), and has an arbitrary value $a_0$ in place of all other existential variables (Line~\ref{line:qpfree-existential-variables}).
As for the annotations, we annotate all facts with $\one$ (Line~\ref{line:non-carry-annotations}) except for those of $R$, which are annotated the same as their source facts.
Therefore, $(D,\tau)$ is $R$-annotated.
See \Cref{fig:q-plus-database-construction} for an illustration of this construction. 

\begin{figure}[t]
  \input{ext-db.pspdftex}
  \caption{An example for the construction from \Cref{lemma:existential-removal-idempotent-negative} for $R$-annotated databases on the query $Q(x_1, x_2) \datarule R(w_1, w_2),S(w_2,x_1),T(x_1,x_2,w_3),U(x_2,w_4,w_5)$. Here, $x_1$ is the carrying variable, $R\carry=\pfreeQ{S}$, and $V=\{w_1,w_2\}.$ \label{fig:q-plus-database-construction}}
\end{figure}

We claim next that the construction time is $O(|D|')$. Indeed, the Yannakakis algorithm requires linear time (using perfect hash tables in the RAM model of computation). Then, Line~\ref{line:iterate-q} iterates over a constant number of atoms, Line~\ref{line:iterate-over-pfree} iterates over a linear number of facts, and Line~\ref{line:iterate-over-atom-variables} iterates over a constant number of variables. Overall, we have a linear number of operations, each requiring constant time. It is left to prove the correctness.

First, we show that every answer tuple $\vec a$ in $Q(D)$ is also in $\Qpfree(D')$.
We claim that the mapping from the free variables to the domain given by $\vec a$ is a homomorphism from $\Qpfree$ to $D'$. 
Consider an atom $\varphi$ of $Q$, and denote its relation symbol by $S$.
First, consider the case that $\varphi$ does not contain a variable in $V$.
Since $D$ contains some $S$-fact that supports $\vec a$ and each $S$-fact was created from an $\pfreeQ{S}$-fact in $D'$ that agrees with it on the free variables, that means that $D'$ contains an $\pfreeQ{S}$-fact that supports $\vec a$.
The second case is that $\varphi$ contains a variable in $V$. That means that every free variable in $\varphi$ is a carrying variable that appears also in $R\carry$. In this case, each $S$-fact was created from an $R\carry$-fact that agrees with it on the free variables of $S$.
As we removed dangling tuples, there must also be an $\pfreeQ{S}$-fact in $D'$ that agrees with that $R\carry$-fact on the free variables of $S$. This $\pfreeQ{S}$-fact supports $\vec a$.

Next, we show that every answer tuple $\vec a$ in $\Qpfree(D')$ is also in $Q(D)$.
Since $\vec a$ is an answer to $\Qpfree(D')$, there exists in $D'$ an $R\carry$-fact $f\carry$ that supports $\vec a$.
We claim that there is a homomorphism from $Q$ to $D$ that assigns the free variables according to $\vec{a}$, assigns $\langle f\carry \rangle$ to the variables of $V$, and assigns $a_0$ to all other existential variables. 
Indeed, consider some atom $\varphi$ of $Q$. If $\varphi$ contains a variable of $V$, then by construction its relation contains a fact that: agrees with $f\carry$ on the carrying variables, has $\langle f\carry \rangle$ in place of any existential variable in $V$, and has $a_0$ in place of any other existential variable.
If the atom $\varphi$ does not contain a variable of $V$, we rely on the fact that the relation of $\pfreeQ{\varphi}$ in $D'$ contains some fact $f'$ that supports $\vec a$ as an answer. We added a fact to the relation of $\varphi$ in $D$ that agrees with $f'$ on the free variables and has $a_0$ in place of any existential variables.
We conclude that the claimed homomorphism exists, and $\vec a$ is also an answer for $Q(D)$.

It is left to show that, given an answer tuple $\vec{a}$ in $\Qpfree(D')$ and $Q(D)$, we have that $\tau(\vec{a})=\tau'(\vec{a})$. 
As $\Qpfree$ is full, there is a single $R\carry$-fact $f\carry$ in $D'$ that supports $\vec a$, and since $\Qpfree$ is $R\carry$ annotated, we know that $\tau'(\vec a) = \tau'(f\carry)$.
In our construction, existential variables not in $V$ are always assigned the value $a_0$, and existential variables in $V$ are always assigned the value $\langle f\carry \rangle$, where $f\carry$ is a fact of $R\carry$. Given an answer $\vec{a}$ to $Q$, there is exactly one homomorphism from $Q$ to $D$ that supports it, the one where free variables are assigned according to $\vec{a}$, the $V$ variables are assigned $\langle f\carry \rangle$, where $f\carry$ is the fact of $R\carry$ that supports $\vec{a}$, and the other existential variables are assigned $a_0$. Since $D$ is locally annotated, $\tau(\vec{a})=\tau(f_R)$, where $f_R$ is the fact supporting $\vec{a}$ in $R$.
If $R$ contains a variable of $V$, then its source is $R\carry$, and $\tau(f_R)=\tau'(f\carry)$. Otherwise, its source is $\pfreeQ{R}$. In this case, since $R$ does not contain any variable of $V$, it means that $V$ is empty by its definition, $R$ contains only free variables, and the carrying variables are exactly the variables of $R$, so $R\carry$ and $\pfreeQ{R}$ have the same variables.
In this case, if $f'$ is the fact from $\pfreeQ{R}$ that constructs $f_R$, we have that $f\carry$ is the fact agreeing with $f'$ in $R\carry$, and our construction sets $\tau(f_R)=\tau'(f\carry)$.
Overall, we have that $\tau(\vec{a})=\tau(f_R)=\tau'(f\carry)=\tau'(\vec{a})$.
\end{proof}

We now have a classification of the \cqstars without self-joins over locally-annotated databases in the case of an idempotent addition.
By \Cref{lemma:existential-removal-idempotent-positive} and \Cref{lemma:existential-removal-idempotent-negative}, direct access for $Q$ over $R$-annotated $\K$-databases is in $\linlog$ if and only if direct access for $\Qpfree$ over $R\carry$-annotated $\K$-databases is in $\linlog$. Then, we only need to apply \Cref{theorem:full-query-annotation-to-variable} on the full $\Qpfree$ to determine the complexity of direct access for $\Qpfree$ based on the structure of its $R\carry$-deannotation.

\begin{thm}\label{theorem:idempotent-r-deannotation}
Let $(\K,\oplus,\otimes,\zero,\one)$ be a logarithmic-time $\oplus$-idempotent commutative semiring.
Let $R$ be a relation symbol of a \cqstar $Q$ without self-joins, and let $Q_R$ be the $R\carry$-deannotation of $\Qpfree$.
\begin{enumerate}
\item If $Q_R$ is acyclic and has no disruptive trio, then direct access for $Q$ is in $\linlog$ on $R$-annotated $\K$-databases.
\item Otherwise, if $\N \subseteq \K$, then direct access for $Q$ is not in $\linlog$ on $R$-annotated $\K$-databases, assuming the \HYPERCLIQUE and \sparseBMM hypotheses. 
\end{enumerate}
\end{thm}

Notice that we can use the positive side of this dichotomy even if $Q$ has self-joins by first assigning different atoms with different relation symbols and copying the relevant relations.
We can use \Cref{theorem:idempotent-r-deannotation} to find tractable cases for \ACQs that have a translation to a $\oplus$-idempotent semiring.

\begin{cor}\label{cor:min-max-tractability}
Consider an \ACQ
$Q(\vec{x}, \alpha(w),\vec{z}) \datarule  \varphi_1(\vec{x},\vec{y},\vec{z}),\dots,\varphi_\ell(\vec{x},\vec{y},\vec{z})$
where either $\alpha$ is $\aggmin$ or $\aggmax$ or $\alpha$ is $\aggcountd$ and $w$ has a logarithmic domain.
Let $Q'(\vec{x}, \star,\vec{z})$ be a \cqstar with the same body as $Q$, and let $Q_R$ be the $R\carry$-deannotation of $Q'_{|\free{(Q')}}$, where $R$ is an atom containing $w$.
If $Q_R$ is acyclic and has no disruptive trio, then direct access for $Q$ is in $\linlog$. 
\end{cor}

\section{Concluding Remarks}\label{sec:conclusions}
Direct access provides an opportunity to efficiently evaluate queries with ordering, even when the number of answers is enormous. Past research studied the feasibility of direct access for CQs, and here we embarked on the exploration of this problem for queries that involve grouping and aggregation, either as standard (SQL) aggregate functions (\ACQs) or annotation with commutative semirings (\cqstars). For \cqstars (and \ACQs that can be efficiently simulated by \cqstars), we showed that the past classification holds as long as the computed value is not involved in the order. This result also holds for count-distinct, but under the more restricted condition that the counted attribute is treated as an ordinary free variable (coming last in the lexicographic order).
Involving the computed value in the order introduces hardness pretty quickly, and we established the narrower tractability condition in this case. We also showed how the condition generalizes when restricting our attention to databases locally annotated by a semiring with an idempotent addition. 

An important direction for future work is the exploration of queries beyond free-connex ones; for that, we need to allow for broader yardsticks of efficiency, as done for direct access for CQs without aggregation~\cite{bringmann2025tight} and as done for Functional Aggregate Queries (FAQ) for CQs with aggregation and traditional query evaluation~\cite{DBLP:conf/pods/KhamisNR16}. Moreover, we plan to explore the extension of our results to queries with self-joins, and we believe that recent results~\cite{bringmann2025tight} can be used towards such an extension.  It is also left for future work to better understand the limits of computation and establish lower bounds (and dichotomies) for general classes of queries, commutative semirings, and aggregate functions. 
Another important direction is to explore the ability to efficiently maintain the direct-access structure for \cqstars and \ACQs through updates of the database, as previously studied in the context of 
aggregate queries~\cite{DBLP:phd/dnb/Keppeler20}
as well as non-aggregate queries~\cite{DBLP:journals/tods/BerkholzKS18,DBLP:journals/tods/SchwentickVZ18}.
Finally, we plan to investigate the practical behavior of our algorithms and understand how well the theoretical acceleration is realized in comparison to existing query engines. 

\section*{Acknowledgment}
  \noindent The work of Idan Eldar and Benny Kimelfeld was supported by the German Research Foundation (DFG) grant KI 2348/1-1 (DIP Program).

\bibliography{sources}
\bibliographystyle{alphaurl}

\end{document}